\documentclass[useAMS,usenatbib]{mn2e}
\input psfig.sty

\def \aj {AJ}
\def \apj {ApJ}
\def \apjl {ApJL}
\def \mnras {MNRAS}
\def \etal {et~al.~}

\def \spose#1{\hbox  to 0pt{#1\hss}}  
\def \lta{\mathrel{\spose{\lower 3pt\hbox{$\sim$}}\raise  2.0pt\hbox{$<$}}}
\def \gta{\mathrel{\spose{\lower  3pt\hbox{$\sim$}}\raise 2.0pt\hbox{$>$}}}

\def \kmsmpc {\>{\rm km}\,{\rm s}^{-1}\,{\rm Mpc}^{-1}}
\def \kms {\ifmmode  \,\rm km\,s^{-1} \else $\,\rm km\,s^{-1}  $ \fi }
\def \kpc {\ifmmode  {\rm kpc}  \else ${\rm  kpc}$ \fi  }  
\def \hMpc {\ifmmode  {\rm h^{-1}Mpc}  \else ${\rm h^{-1}Mpc}$ \fi  }  
\def \Msun {\ifmmode M_{\odot} \else $M_{\odot}$ \fi} 
\def \hMsun {\ifmmode h^{-1}\,\rm M_{\odot} \else $h^{-1}\,\rm M_{\odot}$ \fi}
\def \hhMsun {\ifmmode h^{-2}\,\rm M_{\odot}\else $h^{-2}\,\rm M_{\odot}$ \fi}
\def \Lsun {\ifmmode L_{\odot} \else $L_{\odot}$ \fi} 
\def \hhLsun {\ifmmode h^{-2}\,\rm L_{\odot} \else $h^{-2}\,\rm L_{\odot}$ \fi}

\def\LCDM{$\Lambda$CDM }
\def \LCDM {\ifmmode \Lambda{\rm CDM} \else $\Lambda{\rm CDM}$ \fi}
\def \sig8 {\ifmmode \sigma_8 \else $\sigma_8$ \fi} 
\def \OmegaM {\ifmmode \Omega_{\rm M} \else $\Omega_{\rm M}$ \fi} 
\def \OmegaL {\ifmmode \Omega_{\rm \Lambda} \else $\Omega_{\rm \Lambda}$\fi} 
\def \Deltavir {\ifmmode \Delta_{\rm vir} \else $\Delta_{\rm vir}$ \fi}

\def \rs {\ifmmode r_{\rm s} \else $r_{\rm s}$ \fi} 
\def \Rvir {\ifmmode R_{\rm vir} \else $R_{\rm vir}$ \fi}
\def \Vvir {\ifmmode V_{\rm  vir} \else  $V_{\rm vir}$  \fi} 
\def \Mvir {\ifmmode M_{\rm  vir} \else $M_{\rm  vir}$ \fi}  
\def \Nvir {\ifmmode N_{\rm  vir} \else $N_{\rm  vir}$ \fi}  
\def \Jvir {\ifmmode J_{\rm vir} \else $J_{\rm vir}$ \fi} 
\def \Evir {\ifmmode E_{\rm vir} \else $E_{\rm vir}$ \fi} 
\def \lam {\ifmmode \lambda  \else $\lambda$ \fi} 
\def \lamp {\ifmmode \lambda^{\prime} \else $\lambda^{\prime}$  \fi} 
\def \lampc {\ifmmode \lambda^{\prime}_{\rm c} \else
  $\lambda^{\prime}_{\rm c}$  \fi} 

\def \xoff {\ifmmode x_{\rm off} \else $x_{\rm off}$ \fi}
\def \rhorms {\ifmmode \rho_{\rm rms} \else $\rho_{\rm rms}$ \fi}
\def \qbar {\ifmmode \bar{q} \else $\bar{q}$ \fi}

\def \mgal {\ifmmode m_{\rm gal} \else $m_{\rm gal}$ \fi} 

\def \YB {\ifmmode \Upsilon_B \else $\Upsilon_B$ \fi} 
\def \DeltaIMF {\ifmmode \Delta_{\rm IMF} \else $\Delta_{\rm IMF}$ \fi}

\title[Concentration, Spin and Shape of Dark Matter Haloes]
      {Concentration, Spin and Shape of Dark Matter Haloes:
      Scatter and the Dependence on Mass and Environment}
    
\author[A.V. Macci\`o et al.]    
       {Andrea V. Macci\`o$^{1,3}$\thanks{andrea@physik.unizh.ch}, 
        Aaron A. Dutton$^2$, Frank C. van den Bosch$^3$, Ben Moore$^1$,
       \newauthor{Doug Potter$^1$ \& Joachim Stadel$^1$}\\
      \\
      $^1$Institute for Theoretical Physics, University of Z\"urich,
          Winterthurerstrasse 190 ,CH-8057 Z\"urich, Switzerland \\
      $^2$Department of Physics, Swiss Federal Institute of Technology
          (ETH Z\"urich), CH-8093 Z\"urich, Switzerland \\
      $^3$Max-Planck-Institut f\"ur Astronomie, K\"onigstuhl 17, 69117
          Heidelberg, Germany}

\begin{document}
             
\date{submitted to MNRAS}
             
\pagerange{\pageref{firstpage}--\pageref{lastpage}}\pubyear{2006}

\maketitle           

\label{firstpage}
             

\begin{abstract}
  We  use a  series  of  cosmological N-body  simulations  for a  flat
  $\Lambda$CDM cosmology  to investigate the  structural properties of
  dark matter  haloes, at redshift zero,  in the mass  range $ 3\times
  10^{9}  \hMsun  \lta M_{\rm  vir}  \lta  3\times 10^{13}  $$\hMsun$.
  These properties include the  concentration parameter, $c$, the spin
  parameter,  $\lambda$, and the  mean axis  ratio, $\qbar$.   For the
  concentration-mass  relation   we  find  $c\propto\Mvir^{-0.11}$  in
  agreement  with the  model  proposed by  Bullock  \etal (2001),  but
  inconsistent with the alternative model of Eke, Navarro \& Steinmetz
  (2001).   The  normalization  of  the  concentration-mass  relation,
  however, is 15 percent lower than suggested by Bullock \etal (2001).
  The results  for $\lambda$  and $\qbar$ are  in good  agreement with
  previous studies, when extrapolated  to the lower halo masses probed
  here, while $c$ and $\lambda$ are anti-correlated, in that high-spin
  haloes  have, on average,  lower concentrations.   In an  attempt to
  remove unrelaxed  haloes from the  sample, we compute for  each halo
  the offset  parameter, $\xoff$, defined as the  distance between the
  most bound particle  and the center of mass, in  units of the virial
  radius.   Removing  haloes with  large  $\xoff$  increases the  mean
  concentration by  $\sim 10$ percent, lowers the  mean spin parameter
  by  $\sim 15$  percent, and  removes  the most  prolate haloes.   In
  addition, it  largely removes  the anti-correlation between  $c$ and
  $\lambda$, though  not entirely.   We also investigate  the relation
  between halo  properties and their  large-scale environment density.
  For low  mass haloes we find  that more concentrated  haloes live in
  denser environments than their less concentrated counterparts of the
  same  mass, consistent  with recent  correlation  function analyses.
  Note, however, that the trend  is weak compared to the scatter.  For
  the halo  spin parameters we  find no environment  dependence, while
  there is a weak indication  that the most spherical haloes reside in
  slightly  denser environments.   Finally, using  a simple  model for
  disk  galaxy formation  we show  that haloes  that host  low surface
  brightness galaxies are expected to be hosted by a biased sub-set of
  haloes.   Not only  do these  haloes have  spin parameters  that are
  larger than  average, they  also have concentration  parameters that
  are $\sim 15$  percent lower than the average at  a given halo mass.
  We discuss  the implications of  all these findings for  the claimed
  disagreement between halo  concentrations inferred from LSB rotation
  curves, and those expected for a $\Lambda$CDM cosmology.
\end{abstract}


\setcounter{footnote}{1}


\section{Introduction}
\label{sec:intro}

The theory of  cold dark matter (CDM) provides  a successful framework
for understanding  structure formation  in the universe.   Within this
paradigm dark matter collapses first  into small haloes which merge to
form progressively  larger haloes over time.  Galaxies  are thought to
form out of gas which cools and collapses to the centers of these dark
matter  haloes  (White  \&  Rees  1978).
 
In the  standard picture of  disk galaxy formation the  structural and
dynamical  properties of  disk galaxies  are expected  to  be strongly
related to the properties of the  dark matter haloes in which they are
embedded.   In  particular   the  characteristic  sizes  and  rotation
velocities of  disk galaxies  are determined (to  first order)  by the
spin  parameter, concentration parameter,  and size  of the  host dark
matter   halo  (e.g.   Mo,   Mao  \&   White  1998,   hereafter  MMW).
Consequently, the detailed rotation curve shapes of disk galaxies can,
in principle, be used to  constrain the structural properties of their
dark matter haloes. This is especially true for low surface brightness
(LSB) galaxies, which are believed to be dark matter dominated even at
small radii.  A steadily increasing data base of observed LSB rotation
curves has resulted in a heated debate as to whether the slopes of the
inner  density profiles  of dark  matter dominated  disk  galaxies are
consistent  with the cuspy  profiles found  in N-body  simulations, or
similarly,  whether  the  inferred   concentrations  are  as  high  as
predicted (see Swaters \etal 2003 and references therein).

Unfortunately, determining cusp slopes and/or concentration parameters
from mass modeling rotation curves is non-unique, even for dark matter
dominated  galaxies   (e.g.,  Dutton  \etal   2005).   In  particular,
determining $c$ requires knowledge of the virial radius, which is hard
to constrain  using data that only  covers the inner  $\simeq 10\%$ of
the halo. As an alternative measure  of the central density of a halo,
Alam, Bullock  \& Weinberg (2002) introduced  a dimensionless quantity
that  does  not require  knowledge  of  the  halo virial  radius,  and
demonstrated  convincingly that  the observed  rotation curves  of LSB
galaxies imply halo concentrations  that are systematically lower than
predicted  for a  flat $\Lambda$CDM  cosmology with  a  matter density
$\OmegaM=0.3$ and  a scale-invariant Harrison-Zeldovich power-spectrum
with normalization $\sigma_8=1.0$.

Further  observational  support  for  a  lower  normalization  of  the
$c-\Mvir$  relation  comes  from   the  zero  point  of  the  rotation
velocity-luminosity relation, also  known as the Tully-Fisher relation
(Tully  \&  Fisher  1977)  of  disk galaxies  (van  den  Bosch  2000).
Detailed disk formation models have clearly demonstrated that the high
concentrations of  CDM haloes cause an overprediction  of the rotation
velocities  at a  fixed disk  luminosity,  at least  for a  `standard'
$\LCDM$    cosmology    with    $\OmegaM=0.3$,   $\OmegaL=0.7$,    and
$\sigma_8=0.9$ (e.g., Dutton \etal 2007; Gnedin \etal 2006).

Although these  discrepancies may indicate  a genuine problem  for the
CDM paradigm, there are a number of alternative explanations: First of
all, as shown by various authors (e.g.  Swaters \etal 2003; Rhee \etal
2004;  Spekkens, Giovanelli, \&  Haynes  2005)  the observed  rotation
curves could be hampered by a variety of observational biases, such as
beam smearing, slit offsets and inclination effects, all of which tend
to underestimate the circular velocity in the central regions.

Secondly,  the dark matter  distribution could  have been  modified by
astrophysical  processes   such  as  bars   (e.g.   Holley-Bockelmann,
Weinberg, \& Katz  2005) or dynamical friction (e.g.   Mo \& Mao 2004;
Tonini \etal 2006). These processes tend to lower the concentration of
the  dark  matter halo,  bringing  it  in  better agreement  with  the
observations.   On  the   other  hand,  adiabatic  contraction  (e.g.,
Blumenthal \etal 1986) thought to  be associated with the formation of
disk galaxies, actually tends  to increase the halo concentration, and
it remains to be seen whether the above mentioned processes are strong
enough to undo  this contraction and still cause  a relative expansion
of the inner halo (see Dutton \etal 2007 for a detailed discussion).

A third option is that the data-model comparison has been made for the
wrong  cosmology.   In  particular,   a  reduction  in  the  power  of
cosmological density fluctuations on small scales causes a significant
reduction of the predicted  halo concentrations (e.g.  Eke, Navarro \&
Steinmetz  2001; Zentner  \& Bullock  2002; Alam  \etal 2002;  van den
Bosch, Mo, \& Yang 2003).  Most data-model comparisons have been based
on   a   flat    $\Lambda$CDM   cosmology   with   $\OmegaM=0.3$   and
$\sigma_8=0.9$.  However,  recently the  third year data  release from
the WMAP mission has advocated  a model with $\OmegaM \simeq 0.25$ and
$\sigma_8 \simeq  0.75$ (Spergel  \etal 2006).  This  relatively small
change in  cosmological parameters  causes a significant  reduction of
the  predicted halo  concentration parameters,  bringing them  in much
better agreement with the data (e.g., Yang, Mo, \& van den Bosch 2003;
van den Bosch, Mo \& Yang 2003).

Another potentially important cause for the discrepancy are systematic
errors  in the actual  model predictions.  Both Bullock  \etal (2001a;
hereafter  B01) and Eke,  Navarro \&  Steinmetz (2001;  hereafter ENS)
presented analytical  models that allow  one to compute the  mean halo
concentration   for  given   halo  mass,   redshift  and   cosmology.  
Unfortunately, at  redshift zero the  predictions of these  models are
divergent below  $\sim 10^{11} \Msun$,  with the ENS  model predicting
halo  concentrations  that  are   significantly  lower.   This  is  of
particular importance  for LSB (and  dwarf) galaxies with $V  \lta 100
\kms$, which  are thought  to typically reside  in haloes  with masses
below this  value. Both  B01 and ENS  calibrated their  models against
numerical  simulations. Those  of B01  probed the  mass  range between
$3\times  10^{11} h^{-1}  \Msun$ and  $6\times 10^{13}  h^{-1} \Msun$,
while those of ENS probed an even narrower range from $3\times 10^{11}
h^{-1}  \Msun$ to  $3\times10^{12} h^{-1}  \Msun$ (albeit  with higher
resolution).  What is needed  to discriminate between these models are
simulations that resolve a large  population of low mass haloes, which
is one of the main objectives of this paper.

Another important issue  that we wish to address in  this paper is the
possibility  that  the  LSB  disk  galaxies that  have  been  used  to
constrain  halo concentrations reside  in a  biased sub-set  of haloes
(see discussion  in Wechsler \etal 2006).   Numerical simulations have
shown that there is a  significant scatter in both halo concentration,
$c$, and  halo spin parameter, $\lambda$,  at a given  halo mass (e.g.
Bullock \etal 2001a,b).  Thus if disk galaxies form in a biased subset
of haloes, this  could lead to an apparent  discrepancy between theory
and observation. In  fact, there are a number  of potential causes for
such  a   bias.   First  of   all,  disk  galaxies  are   expected  to
preferentially  form in haloes  that have  not experienced  any recent
major  merger.  There is  evidence that  such a  subset of  haloes has
higher mean $c$,  lower mean $\lambda$, and lower  scatter in both $c$
and  $\lambda$  (Wechsler  \etal  2002;  D'Onghia  \&  Burkert  2004).
Clearly, this would worsen the disagreement between model and data. On
the  other  hand,  it  has  also  been  suggested  that  LSB  galaxies
preferentially  reside in haloes  with relatively  low concentrations.
First  of  all,   since  disks  are  thought  to   be  in  centrifugal
equilibrium,  less concentrated haloes  will harbor  less concentrated
(i.e., lower  surface brightness) disk galaxies (e.g.,  MMW, B01).  In
addition, using  numerical simulations Bailin \etal  (2005) found that
haloes   with  higher   spin  parameters   have,  on   average,  lower
concentration parameters.  Since LSB  galaxies are thought to be those
with  high spin  parameters such  a correlation  would imply  that LSB
galaxies  reside in  haloes  with relatively  low concentrations.   If
confirmed this could offer an alternative explanation as to why (some)
LSB galaxies have lower concentrations than predicted.  Note, however,
that previous  studies (B01, Navarro,  Frenk \& White  1997, hereafter
NFW),   have  found   no  correlation   between  spin   parameter  and
concentration.

Another potential bias for disk  galaxy formation could arise if there
is  a correlation  between  environment (defined  as  the large  scale
matter  density) and $c$  or $\lambda$.   In particular,  Harker \etal
(2006) found  evidence  that low  mass  haloes  in dense  environments
assemble  earlier  than  haloes   of  the  same  mass  in  under-dense
environments. Note  however, that for the  lowest density environments
this trend reverses so that formation redshifts actually increase with
decreasing  density.   Since  haloes  that  assemble  later  are  less
concentrated  (Wechsler \etal  2002), one  thus may  expect  a similar
correlation between halo  concentration and environmental density.  If
dwarf and LSB galaxies preferentially form in under-dense regions this
could also help explain the lower than expected halo concentrations of
these galaxies.

In this  paper we study galaxy size  dark matter haloes from  a set of
cosmological N-body simulations with  the following goals: (i) to test
the predictions  of B01 and  ENS regarding the halo  concentrations of
low mass haloes  (down to $\sim 3 \times 10^9  h^{-1} \Msun$), (ii) to
determine the scatter in concentration, spin parameter, and halo shape
at a  given mass,  (iii) to determine  whether there is  a correlation
between the  spin and concentration parameters, and  (iv) to determine
whether $c$,  $\lambda$ and  halo shape depend  on the density  of the
environment in which  the halo is located.  Our  paper is organized as
follows.   Section~2  describes our  set  of  N-body simulations.   In
Section~3 we  discuss how halo concentration, halo  spin parameter and
halo shape  depend on halo mass.  Sections~4  investigates whether $c$
and  $\lambda$  are  correlated,   while  Section~5  focuses  on  the
environment dependence of halo  properties. In Section~6 we use simple
models  for disk  formation  to investigate  whether  one expects  LSB
galaxies to  reside in haloes  with a biased concentration  parameter. 
Finally, we summarize our results in Section~7.

\section{N-body simulations} 
\label{sec:results}

In order to  explore as wide a range of virial  masses as possible, we
run  simulations of  4 different  box sizes,  listed in  Table~1.  For
comparison we also show the parameters of the Bullock \etal (2001a,b),
Bailin  \etal   (2005)  and  millennium  run   (Springel  \etal  2005)
simulations.  The B01 simulation  has similar size and mass resolution
as  our 64$_{a,b}$ boxes,  while  our smallest  box simulation  has a  mass
resolution that  is $\sim  5$ times higher  than that of  Bailin \etal
(2005).  In  order to  test for cosmic  variance, and to  increase the
size  of our  sample we  ran  two simulations  for each  of the  three
smallest box sizes.

All simulations have been performed  with PKDGRAV, a tree code written
by Joachim Stadel and Thomas Quinn (Stadel 2001). The code uses spline
kernel softening, for which  the forces become completely Newtonian at
2  softening lengths.   Individual time  steps for  each  particle are
chosen  proportional  to the  square  root  of  the softening  length,
$\epsilon$,    over   the   acceleration,    $a$:   $\Delta    t_i   =
\eta\sqrt{\epsilon/a_i}$. Throughout, we set $\eta = 0.2$, and we keep
the  value of the  softening length  constant in  comoving coordinates
during each run. The physical values of $\epsilon$ at $z=0$ are listed
in Table~1. Forces  are computed using terms up  to hexadecapole order
and  a  node-opening  angle  $\theta$  which  we  change  from  $0.55$
initially to $0.7$ at $z=2$.  This allows a higher force accuracy when
the mass distribution  is nearly smooth and the  relative force errors
can be large.
 
We adopt a flat $\Lambda$CDM  cosmology with parameters from the first
year  WMAP results  (Spergel \etal  2003): matter  density  $\OmegaM =
0.268$, baryon  density $\Omega_b = 0.044$, Hubble  constant $h \equiv
H_0/(100 \kmsmpc)  = 0.71$, and  a scale-invariant, Harrison-Zeldovich
power-spectrum  with  normalization  $\sigma_8  =  0.9$\footnote  {The
recent  analysis of  the three  year  WMAP data  (Spergel \etal  2006)
suggests  lower  values for  $\OmegaM$,  $\sigma_8$  and the  spectral
index.   In a forthcoming  paper (Macci\`o  \etal, in  preparation) we
investigate  the effects of  these new  cosmological parameter  on our
results.   The  main  change  regards  a lower  normalization  of  the
concentration, as expected from the B01 and ENS models.}.  The initial
conditions are generated with the GRAFIC2 package (Bertschinger 2001),
which  also computes  the  transfer  function as  described  in Ma  \&
Bertschinger (1995).  The starting redshifts $z_i$ are set to the time
when  the  standard deviation  of  the  smallest density  fluctuations
resolved within  the simulation box reaches $0.2$  (the smallest scale
resolved  within  the  initial  conditions  is defined  as  twice  the
intra-particle distance,  while the  maximum scale is  set by  the Box
size).

\begin{table}
 \centering
 \begin{minipage}{140mm}
  \caption{N-body Simulation Parameters}
  \begin{tabular}{lcccc}
\hline  Name &  Box  size  & N  &  particle mass  &  force  soft. \\  
& $[h^{-1}{\rm Mpc}]$ &  & $[h^{-1}M_{\odot}]$  & $[h^{-1}{\rm kpc}]$ \\ 
14a, 14b & 14.2  & $250^3$ & $1.4\times 10^7$ & 0.43 \\ 
28a, 28b & 28.4  & $250^3$ & $1.1\times 10^8$ & 0.85 \\ 
64a, 64b & 63.9  & $300^3$ & $7.2\times 10^8$ & 1.92 \\ 
128      & 127.8 & $300^3$ & $5.8\times 10^9$ & 3.83 \\ 
\hline 
Bullock    & 60    & $256^3$ & $1.1\times 10^9$ & 1.8 \\ 
Bailin     & 50    & $512^3$ & $7.8\times 10^7$ & 5.0 \\ 
Millennium & 500   & $2160^3$& $8.6\times 10^8$ & 5.0 \\ 
\hline

\end{tabular}
\end{minipage}
\end{table}

In all of our numerical  simulations, haloes are identified using a SO
(Spherical Overdensity) algorithm.  As  a first step, candidate haloes
are  located using  the standard  friends-of-friends  (FOF) algorithm,
with a linking length $b n^{-1/3}$, with $n$ the mean particle density
and $b$  a free  parameter which we  set to  $0.2$.  We only  keep FOF
haloes with at least $N_{\rm  min}=200$ particles, which we subject to
the following  two operations: (i) we  find the point,  $C$, where the
gravitational potential due to the  group of particles is minimum, and
(ii) we determine the radius $\Rvir$, centered on $C$, inside of which
the  density  contrast  is  $\Deltavir$.  For  our  adopted  cosmology
$\Deltavir  \simeq 98$ (using  the fitting  function of  Mainini \etal
2003).  Using all particles in the corresponding sphere we iterate the
above procedure until we converge onto a stable particle set.  The set
is discarded if, at some  stage, the sphere contains less than $N_{\rm
min}$ particles.  If a particle is a potential member of two haloes it
is assigned to the more massive  one.  For each stable particle set we
obtain the virial radius, $\Rvir$,  the number of particles within the
virial radius,  $\Nvir$, and  the virial mass,  $\Mvir$. Above  a mass
threshold of  $\Nvir = 250$  particles there are $\sim  2750$, $3750$,
$7450$ and  $4500$ haloes in the  simulations of box  size 14.2, 28.4,
63.9, 127.8  $h^{-1} {\rm Mpc}$  respectively (these numbers  refer to
the two versions of each box size combined together).

\begin{figure*}
\psfig{figure=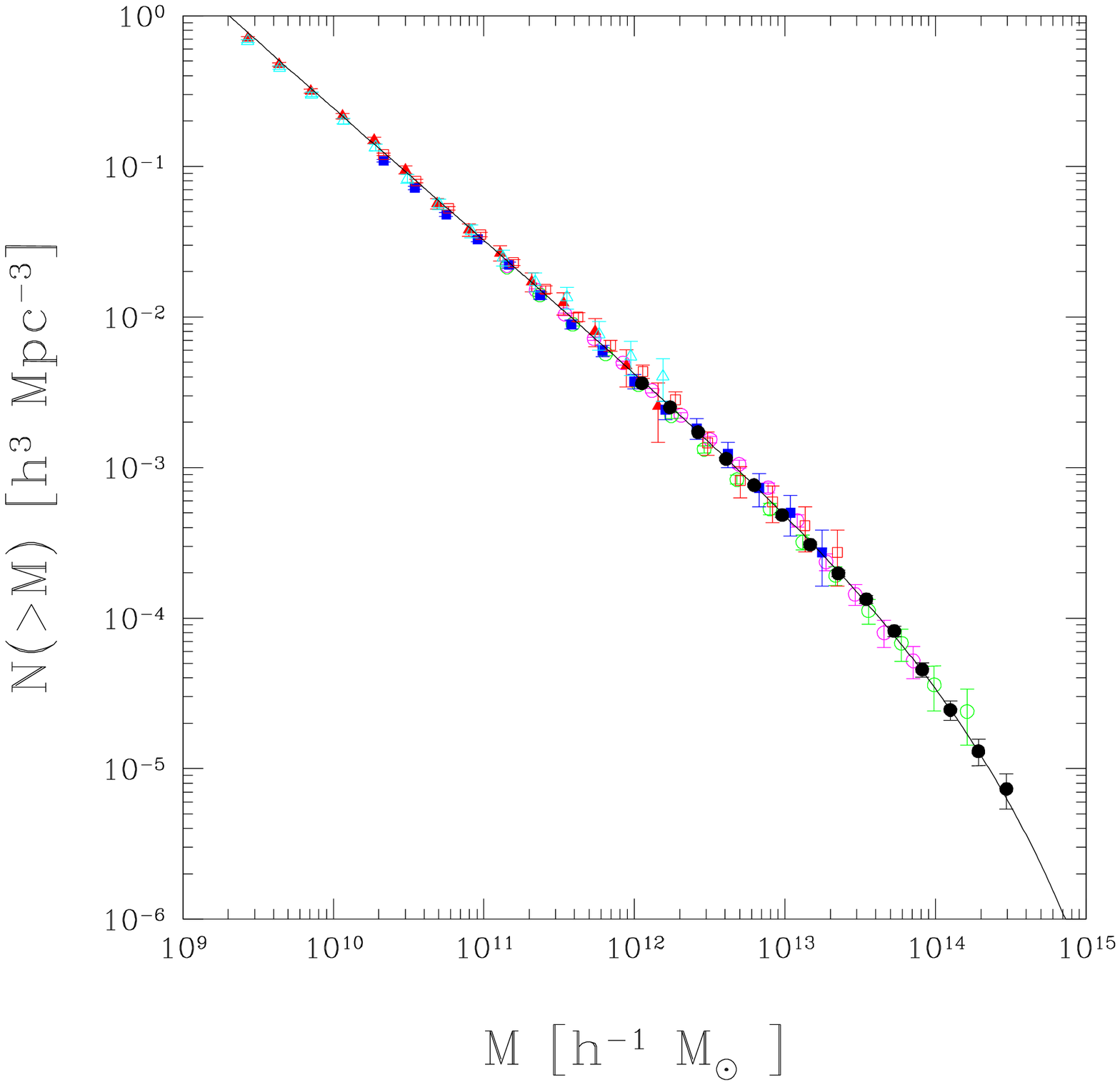,width=500pt}
\caption{Mass functions for the simulations. The point and color types
  correspond to  the different runs: 14a (cyan  triangles); 14b (red
  triangles);  28a  (blue   squares);  28b  (yellow  squares);  64a
  (magenta  open circles);  64b  (green open  circles); 128  (black
  circles).  The solid  line is  the  Sheth \&  Tormen prediction  for
  $\sigma_8=0.9$.}
\label{fig:mf}
\end{figure*}

In  Fig.~ \ref{fig:mf}  we  report  the comparison  of  the halo  mass
functions of all our simulations  with the analytical mass function of
Sheth \&  Tormen (2002). Since the  Sheth \& Tormen  mass function has
been  tuned  to  reproduce  the  mass function  of  FOF  haloes  (with
$b=0.2$), we use  the same FOF masses here. For  the remainder of this
paper,  however, we  consistently will  use the  spherical overdensity
masses, $\Mvir$,  described above.  Note  that the FOF  mass functions
agree well with  the Sheth \& Tormen mass function  over the full five
orders of magnitude  in halo mass probed by  our simulations: all data
points are consistent with the model within one sigma (error bars show
the Poisson  noise in each  bin due to  the finite number  of haloes).
Moreover all the simulations made  with different box sizes agree with
each other in the mass ranges where they overlap.

\subsection{Halo parameters}

For each halo we determine a set of parameters as described below. All
of  these parameters are  derived from  the SO  haloes (i.e.  from the
particle sets  defined by the SO  criteria), rather than  from the FOF
particle sets.

\subsubsection{concentration parameter}

N-body simulations  have shown  that the spherically  averaged density
profiles of  DM haloes can  be well described  by a two  parameter NFW
profile:
\begin{equation}
\frac{\rho(r)}{\rho_{\rm crit}} = \frac{\delta_{\rm c}}{(r/\rs)(1+r/\rs)^2},
\label{eq:nfw}
\end{equation}
where  $\rho_{\rm crit}$  is  the critical  density  of the  universe,
$\delta_{\rm c}$  is the characteristic  overdensity of the  halo, and
$\rs$ is  the radius where the  logarithmic slope of  the halo density
profile  ${\rm d}\ln\rho/{\rm  d}\ln r  =  -2$ (NFW).   A more  useful
parametrization  is  in  terms  of  the  virial  mass,  $\Mvir$,  and
concentration  parameter,  $c\equiv\Rvir/\rs$.   The virial  mass  and
radius are related by $\Mvir = \Delta_{\rm vir} \rho_{\rm crit} (4 \pi
/ 3) \Rvir^3$, where $\Delta_{\rm vir}$ is the density contrast of the
halo.

To compute the concentration of  a halo we first determine its density
profile. The halo center is defined  as the location of the most bound
halo particle, and  we compute the density ($\rho_i$)  in 50 spherical
shells,  spaced equally  in  log  radius. Errors  on  the density  are
computed from the Poisson noise  due to the finite number of particles
in each mass  shell.  The resulting density profile is  fit with a NFW
profile (Eq.~\ref{eq:nfw}),  which provides a good fit  to most haloes
over the  range of  radii we  are interested in.   Note that,  in this
paper, we  are not  concerned with the  inner asymptotic slope  of the
density profile. During the fitting  procedure we treat both $\rs$ and
$\delta_c$   as  free  parameters.    Their  values,   and  associated
uncertainties,  are  obtained via  a  $\chi^2$ minimization  procedure
using the Levenberg  \& Marquart method. We define  the r.m.s.  of the
fit as:
\begin{equation}
\rhorms = \frac{1}{N}\sum_i^N { (\ln \rho_i - \ln \rho_m)^2}
\label{eq:rms}
\end{equation}
where $\rho_{\rm m}$ is the fitted NFW density distribution. We do not
use  the $\chi^2$  value of  the  best-fit since  this increases  with
$\Nvir$.   This occurs  because higher  resolution haloes  have better
resolved substructure and smaller  Poisson errors on the density, thus
making the  fit worse.  Finally,  we compute the concentration  of the
halo, $c$, using the virial radius obtained from the SO algorithm, and
we define the error on $\log c$ as $(\sigma_{\rs}/\rs)/\ln(10)$, where
$\sigma_{\rs}$ is the fitting uncertainty on $\rs$.

We checked our concentration fit pipeline against the one suggested by
B01. As a test  we used both the procedures to compute  $c$ in all our
cubes.  No systematic  offset arises  in the  concentration  {\it vs.}
mass  relation  due  to  the  different halo  definition  and  fitting
procedure.

\subsubsection{spin parameter}
\label{sec:spinpar}

The  spin  parameter is  a  dimensionless  measure  of the  amount  of
rotation of a  dark matter halo.  The standard  definition of the spin
parameter, due to Peebles (1969), is given by
\begin{equation}
\label{eq:lambda}
{\lambda = \frac{\Jvir |\Evir|^{1/2}}{G\Mvir^{5/2}},}
\end{equation}
where  $\Mvir,$  $\Jvir$  and  $\Evir$  are the  mass,  total  angular
momentum  and energy of  the halo,  respectively. Due  to difficulties
with accurately measuring $\Evir$,  Bullock \etal (2001b) introduced a
modified spin parameter:
\begin{equation}
\lamp=\frac{\Jvir}{\sqrt{2}\Mvir\Vvir\Rvir}
\end{equation}
with  $\Vvir$ the  circular  velocity  at the  virial  radius.  For  a
singular isothermal sphere these  two definitions are equivalent.  For
a pure NFW halo, however, they are related according to $\lambda=\lamp
f(c)^{1/2}$ with $f(c)=\frac{1}{2} c  [ (1+c)^2 - 1 -2(1+c)\ln(1+c)] /
[c  -(1+c)\ln(1+c)]^2$  (MMW).  In  what  follows,  we define  $\lampc
\equiv \lamp f(c)$ as the ``corrected'' spin parameter.

In  order  to  avoid  potential  problems and  inaccuracies  with  the
measurement of $\Evir$,  we adopt the $\lamp$ definition  for the halo
spin parameter,  unless specifically stated  otherwise.  The advantage
of $\lamp$ over  $\lampc$ is that the latter  can introduce artificial
correlations between halo concentration and halo spin parameter, since
an error in $c$ translates into an error in $f(c)$ and hence $\lampc$.
We define the uncertainty  in $\log\lamp$ as $(\sigma_J/J) / \ln(10)$,
were we use that $\sigma_J/J = \sqrt{\frac{1}{N}(1 + 0.04/\lam^{\prime
    2})} \simeq 0.2/\lamp \sqrt(N)$  (Bullock \etal 2001b).  Note that
this implies that the errors on  $\lamp$ are largest for haloes with a
low spin parameter and with few particles.

\subsubsection{shape parameter}

Determining the shape of a three-dimensional distribution of particles
is a  non-trivial task  (i.e. Jing \&  Suto 2002).   Following Allgood
\etal (2006)  we determine the shape  of our haloes  starting from the
inertia tensor.   As a first  step the inertia  tensor of the  halo is
computed using all the particles within the virial radius; in this way
we  obtain  a  $3  \times  3$  matrix.  Then  the  inertia  tensor  is
diagonalized and the particle distribution is rotated according to the
eigen  vectors.  In this  new frame  (in which  the moment  of inertia
tensor is diagonal) the  ratios $a_2/a_1$ and $a_3/a_1$ ($a_1,a_2,a_3$
being the major, intermediate  and minor axis, respectively) are given
by:
\begin{equation}
{a_2 \over a_1} = \sqrt{ { \sum m_i y_i^2} \over \sum { m_i x_i^2}},
\newline
{a_3 \over a_1} = \sqrt{ { \sum m_i z_i^2} \over \sum { m_i x_i^2}}.
\end{equation}

Next we again compute the inertia tensor, but this time only using the
particles  inside the  ellipsoid defined  by $a_1$,  $a_2$ and  $a_3$. 
When deforming the ellipsoidal volume of the halo, we keep the longest
axis ($a_1$) equal to the original radius of the spherical volume (cf.
Allgood \etal 2006).  We iterate this procedure until we converge to a
stable set of the axis ratios.

Since dark matter  haloes tend to be prolate,  a useful parameter that
describes the shape of the  halo is $\bar{q} \equiv (a_2 + a_3)/2a_1$,
with  the limiting  cases being  a sphere  ($\bar{q}=1$) and  a needle
($\bar{q}=0$).
 
\subsubsection{offset parameter}

The  last  quantity that  we  compute for  each  halo  is the  offset,
$\xoff$, defined as the distance between the most bound particle (used
as the center  for the density profile) and the center  of mass of the
halo, in units of the virial radius.  This offset is a measure for the
extent to  which the  halo is relaxed:  relaxed haloes  in equilibrium
will have a smooth,  radially symmetric density distribution, and thus
an offset that is virtually  equal to zero.  Unrelaxed haloes, such as
those that have  only recently experienced a major  merger, are likely
to  reveal  a  strongly  asymmetric  mass  distribution,  and  thus  a
relatively large  $\xoff$. Although some  unrelaxed haloes may  have a
small $\xoff$, the advantage of  this parameter over, for example, the
actual  virial  ratio, $2T/V$,  as  a  function  of radius  (Macci\`o,
Murante \&  Bonometto 2003;  Shaw \etal 2006),  is that the  former is
trivial to evaluate.

Fig.~\ref{fig:all5}  shows  histograms  of, and  correlations  between
$\Nvir$, $\rhorms$,  $\xoff$, and $\bar{q}$.   The rms of  the density
profile  fit decreases with  $\Nvir$, as  expected, while  $\xoff$ and
$\qbar$ are  uncorrelated with $\Nvir$  (especially for $\Nvir>10^3)$.
The solid  lines in  the $\xoff$-$\Nvir$ plot  show the ratios  of the
softening length  to the  virial radius.  This  shows that  the offset
parameter is  not effected  by resolution effects.   In an  ideal halo
simulation with  a large  number of particles  $\xoff$ is  expected to
decrease with the decrease of the halo mass.  We do not see this trend
in the  $\xoff$-$\Nvir$ relation because in our  simulations the value
of $\xoff$  is mostly dominated by  numerical effects at  the low mass
tail. Moreover since we used different simulations with different mass
resolution  there  is not  a  one to  one  relation  between mass  and
$\Nvir$.  Note  also that $\xoff$ is uncorrelated  with $\rhorms$, but
that there is a strong correlation between $\xoff$ and $\qbar$ so that
more  prolate haloes  tend to  have  larger offsets.   We discuss  the
implications of this correlation in \S\ref{sec:unrelaxed}.

The  distributions of  $\rhorms$ and  $\log(\xoff)$  are approximately
normal with means of $0.4$ and $-1.4$, respectively.  The distribution
of $\qbar$, on the other hand,  is strongly skewed. Note also that the
$\qbar$-distributions are slightly  different for different box-sizes. 
This  is because there  is a  correlation between  halo mass  and halo
shape,  such  that  more   massive  haloes  are  less  spherical  (see
Section~\ref{sec:unrelaxed} below).   Since larger boxes  contain more
massive haloes, this results in lower mean axis ratios.


\section{Mass dependence of Spin, Concentration and Shape}
\label{sec:massdep}

\begin{figure*}
\psfig{figure=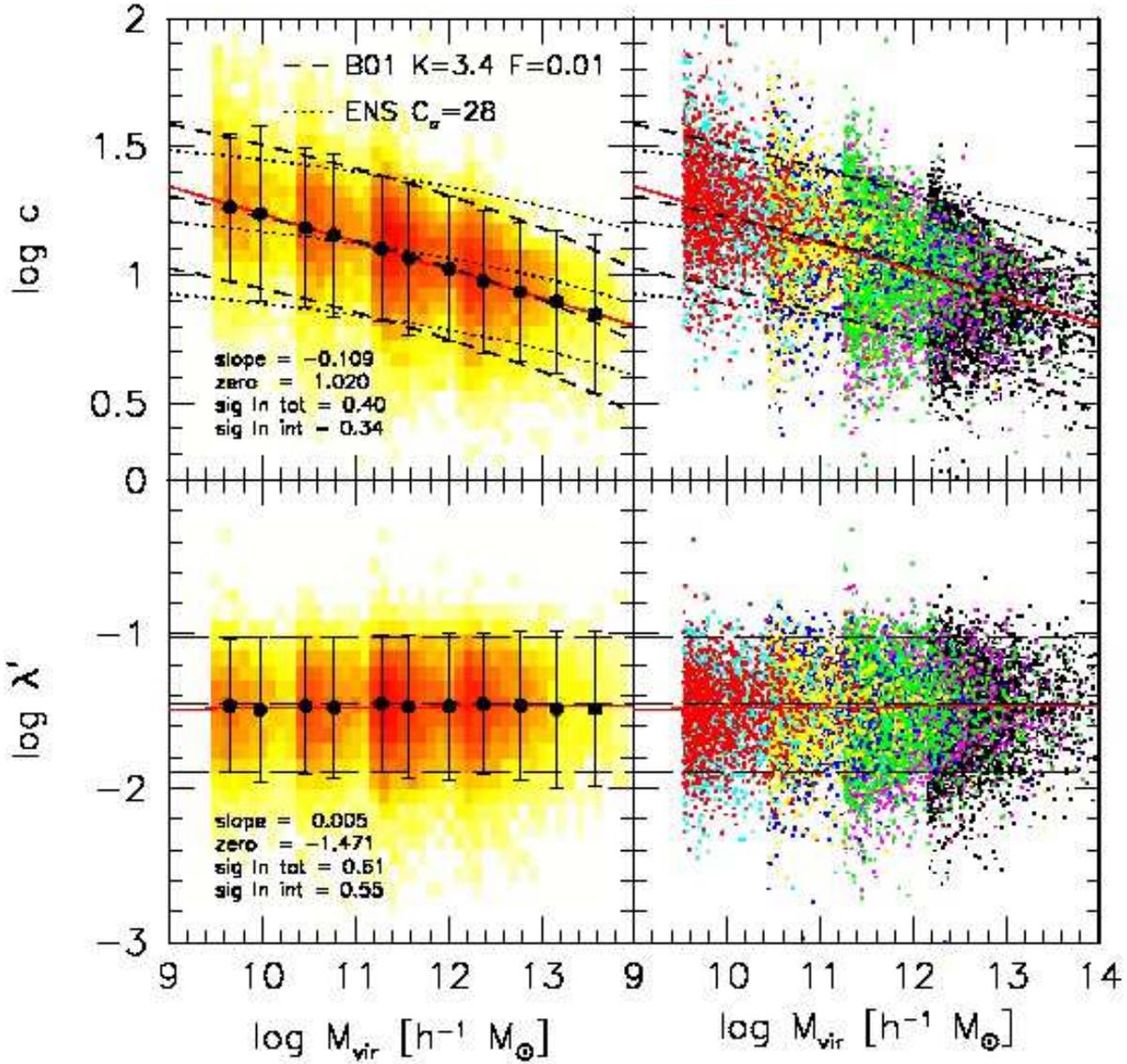,width=500pt}
\caption{Relations between concentration, spin parameter and virial 
  mass for haloes  with $\Nvir > 250$.  The dashed  lines give the mean
  and $2\sigma$ scatter from  Bullock \etal 2001a,b. Dotted lines give
  the  ENS  model  prediction.   The  solid red  lines  give  weighted
  power-law  fits: $y={\rm  zero}  +{\rm slope}(\log\Mvir-12)$,  where
  $y=\log  c$  or  $  \log   \lambda$  and  $\Mvir$  is  in  units  of
  $h^{-1}\Msun$.  The  parameters of the  fits are given in  the lower
  left corner  of each panel.   The panels on  the left show  the data
  binned in  mass.  The filled  circles give the error  weighted mean,
  the small error  bars gives the Poisson error on  the mean while the
  larger error bars give the intrinsic 2-sigma scatter.  The panels on
  the  right show all  the data  points color  coded according  to the
  simulation:  14a (cyan);  14b (red);  28a (blue);  28b (yellow);
  64a  (magenta); 64b  (green); 128  (black).}
\label{fig:clm_all}
\end{figure*}

Fig.~\ref{fig:clm_all} shows the relations  of concentration {\it vs.} 
mass, and spin parameter {\it vs.}  mass for all haloes with more than
250 particles  within the  virial radius.  The  right panels  show the
data from  the simulations  with the four  box sizes clearly  visible. 
The  left panels show  the mean  (solid dots)  and twice  the standard
deviation of  $c$ and $\lamp$ (error  bars) in bins  equally spaced in
$\log\Mvir$ (plotted at  the mean $\Mvir$ in each  bin).  The mean $c$
and  $\lamp$  are  computed  taking  account  of  both  the  estimated
measurement errors and the intrinsic scatter, using
\begin{equation}
\bar{y} = \frac{\Sigma_i y_i w_i}{\Sigma_i w_i},  
\;\;w_i = (\sigma_{\rm int}^2 + \sigma_i^2)^{-1}
\end{equation}
Here $y_i$ denotes  either $c$ or $\lamp$ of the  $i$th halo, $w_i$ is
the weight (haloes with larger uncertainties receive less weight), and
$\sigma_i$ is the  measurement error on $c$ or  $\lamp$. The intrinsic
scatter $\sigma_{\rm int}$ is given by
\begin{equation}
\sigma_{\rm int}^2 = \frac{ \Sigma_i \left[(y_i - \bar{y})^2
  -\sigma_i^2 \right] w_i} {\Sigma_i w_i}.
\end{equation}
Since these two equations are coupled the computation of $\bar{y}$ and
$\sigma_{\rm  int}$  requires an  iterative  procedure.   We start  by
computing  $\bar{y}$  and $\sigma_{\rm  int}$  with  $w_i=1$. Next  we
iterate until  a stable solution for $\bar{y}$  and $\sigma_{\rm int}$
is  found.   This  procedure   converges  rapidly,  typically  in  3-4
iterations.

The  solid red  lines in  Fig.~\ref{fig:clm_all} show  weighted (using
errors on  $c$ and $\lamp$) least  squares fits of $c$  and $\lamp$ on
$\Mvir$.  The $c-\Mvir$ relation is  well fitted by a single power-law
over four orders  of magnitude in mass $3\times  10^{9} \, \hMsun \lta
\Mvir \lta 3\times 10^{13}\, \hMsun$, with
\begin{equation}
\label{fullc}
\log c = 1.020[\pm0.015] -0.109[\pm0.005](\log\Mvir -12).
\end{equation}
Note that $\Mvir$ is in units of $h^{-1}\Msun$.  The numbers in square
brackets give the scatter in the corresponding value between the seven
different simulations.  The total  scatter about this mean relation is
$\sigma_{\ln c}=0.40\pm0.03$ and the intrinsic scatter is $\sigma_{\ln
c}=0.33\pm0.03$,  where  again  the  uncertainty  corresponds  to  the
box-to-box scatter.   These are in excellent agreement  with the total
and  intrinsic scatter  found  by  B01 which  are  $0.41$ and  $0.32$,
respectively  (see  also  Wechsler  \etal  2002).  The  slope  of  the
$\lamp-\Mvir$  relation is  consistent  with zero,  in agreement  with
previous studies  (e.g.  Lemson \&  Kauffmann 1999, hereafter LK99; 
Shaw  \etal 2006).
If we  take $\lamp$  to be  independent of $\Mvir$,  we find  a median
$\lamp=0.034\pm0.001$     with     an     intrinsic     scatter     of
$\sigma_{\ln\lamp}=0.55\pm0.01$, which are in excellent agreement with
Bullock  \etal   (2001b)  (median  of  $\lamp=0.035   \pm  0.005$  and
$\sigma_{\ln \lamp}=0.50  \pm 0.03$) and other studies  (e.g., van den
Bosch \etal  2002; Avila-Reese \etal  2005)\footnote{Haloes with lower
$\lamp$ have,  on average, larger errors,  which results in  a tail of
haloes  with   low  $\lamp$.   As  long  as   the  larger  measurement
uncertainties of  these haloes are  properly taken into  account, this
tail  does   not  effect   the  mean  and   scatter  of   the  $\lamp$
distribution.}.

The dashed and  dotted lines show the mean $c(\Mvir)$  for the B01 and
ENS models, respectively.   In addition, for both models  we also show
the upper and lower $2\sigma$ intrinsic scatter bounds, where we adopt
$\sigma_{\ln  c}=0.32$ (Wechsler \etal  2002). The  B01 model  has two
free  parameters:  $K$,  which  determines the  normalization  of  the
relation, and  $F$, which effects the  slope.  The ENS  model has just
one free parameter $C_{\sigma}$  which controls the normalization; the
slope  is fully determined  by the  model.  If  we adopt  $F=0.01$, as
advocated  by B01,  we find  that the  slope of  the B01  model  is in
excellent agreement with our simulations over the full range of masses
covered.    Consequently,   our   data   is  inconsistent   with   the
significantly  shallower  slope  of  the  ENS model.   Note  that  ENS
compared their model  to a very small sample  of relaxed haloes ($\sim
15$), albeit  with high  resolution ($\Nvir >  30 000$), over  a small
mass range  $\Mvir \simeq 2\times 10^{11}-4\times 10^{12}  \; \hMsun$. 
Over this mass range the ENS model is in reasonable agreement with our
simulation data.

In terms  of the normalization  our data is  best fit with  $K=3.4 \pm
0.1$ (for  $F=0.01$), where  the error reflects  the effect  of cosmic
variance as determined from  the various independent simulation boxes.
This is 15 percent lower than the $K=4.0$ originally advocated by B01,
but consistent  with Zhao  \etal (2003) and  Kuhlen \etal  (2005), who
found $K=3.5$ (both for $F=0.01$).  In their paper Kuhlen \etal argued
that  the cause  for their  lower normalization  might be  due  to the
$N$-body code used for their simulations (GADGET-1).  However, we have
used an  independent code,  PKDGRAV, and obtain  the same  result.  We
therefore  suspect  that  the   cause  for  this  discrepancy  resides
somewhere  else.   Indeed,  as  it  turns out,  B01  used  a  slightly
different transfer  function for setting up the  initial conditions of
their numerical  simulation than for  computing their model.   If they
correct this,  they obtain  a best  fit $K =  3.75$ (James  Bullock \&
Andrew Zentner,  private communication).   If we take  our (admittedly
crude)  estimate for  the cosmic  variance at  face value,  this still
implies that  we predict a  normalization that is  significantly lower
(at the $3\sigma$ level).

As pointed  out by B01, a better  match to the slope  of the $c-\Mvir$
relation for haloes more massive  than $\simeq 10^{13} \hMsun$ (at the
expense of  worsening the  agreement for haloes  with $M  \lta 10^{11}
\hMsun$) can  be obtained  by using $F=0.001$,  in which case  we find
that $K=2.6$.  This is again approximately 15 percent smaller than the
$K=3.0$ advocated by B01 for this value of $F$.


\subsection{The Impact of Unrelaxed Haloes} 
\label{sec:unrelaxed}

Our  halo finder  (and halo  finders  in general)  do not  distinguish
between relaxed  and unrelaxed  haloes. There are  two reasons  why we
might want  to remove unrelaxed  halos.  First, and  most importantly,
unrelaxed haloes  often have poorly  defined centers, which  makes the
determination  of   a  radial  density  profile,  and   hence  of  the
concentration  parameter,   an  ill-defined  problem.    Secondly  for
applications  to  disk  galaxy  formation,  haloes  that  are  not  in
dynamical  equilibrium are unlikely  to host  disk galaxies,  and even
less likely to host the more dynamically fragile LSB galaxies. In this
case,  the halo parameters  inferred from  (LSB) disk  rotation curves
need to be compared to those of the subset of relaxed haloes only.

\begin{figure*}
\psfig{figure=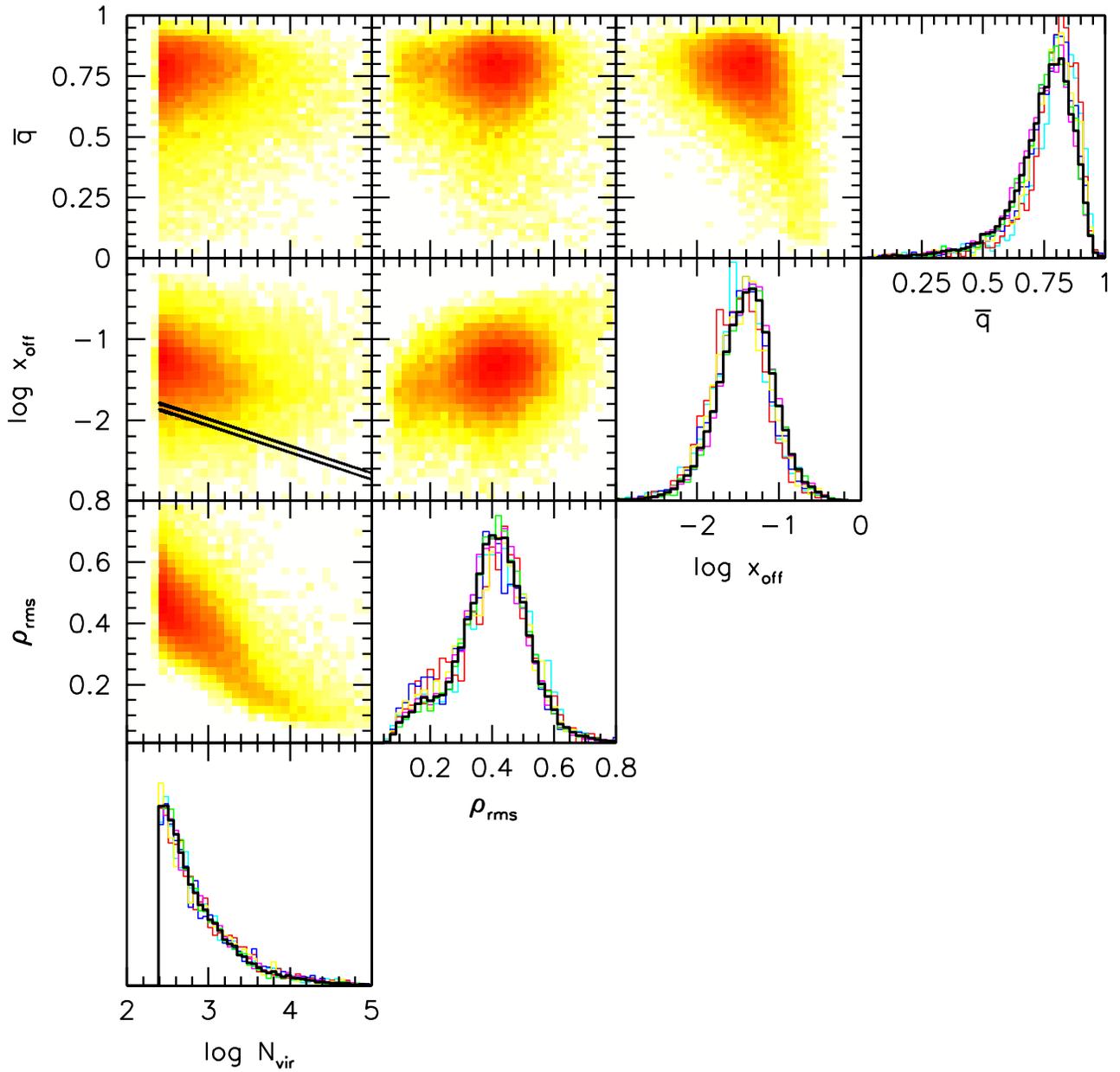,width=500pt}
\caption{Correlations between, $N_{\rm vir}$, $\rho_{\rm rms}$,
  $x_{\rm off}$, and  $\qbar$.  The color coding in  the density plots
  is according to the logarithm of  the number of points in each cell. 
  For  the   histograms  the   colors  correspond  to   the  different
  simulations as in Fig.~\ref{fig:mf},  the thick black line shows the
  histograms  of  the  combined   samples.  The  black  lines  in  the
  $\xoff-\Nvir$ plot  shows the ratio  of the softening length  to the
  virial radius.}
\label{fig:all5}
\end{figure*}

One could imagine  using $\rhorms$ (the r.m.s.  of the  NFW fit to the
density profile) to decide whether a halo is relaxed or not.  However,
while  it is  true  that  $\rhorms$ is  typically  high for  unrelaxed
haloes,  haloes  with  relatively  few  particles  also  have  a  high
$\rhorms$ (due to  Poisson noise) even when they  are relaxed. This is
evident  from  Fig.~\ref{fig:all5},  which  shows that  $\rhorms$  and
$\Nvir$ are strongly correlated.  Furthermore not all unrelaxed haloes
have a high  $\rhorms$.  We found several examples  of haloes with low
$\rhorms$ which are  clearly unrelaxed.  This is due  to the smoothing
effects  of spherical averaging  when computing  the density  profile. 
However, these haloes are often  characterized by a large $\xoff$ (the
offset between  the most  bound particle and  the center of  mass). In
what  follows we  therefore use  both $\rhorms$  and $\xoff$  to judge
whether a halo is relaxed or not.

\begin{figure*}
\psfig{figure=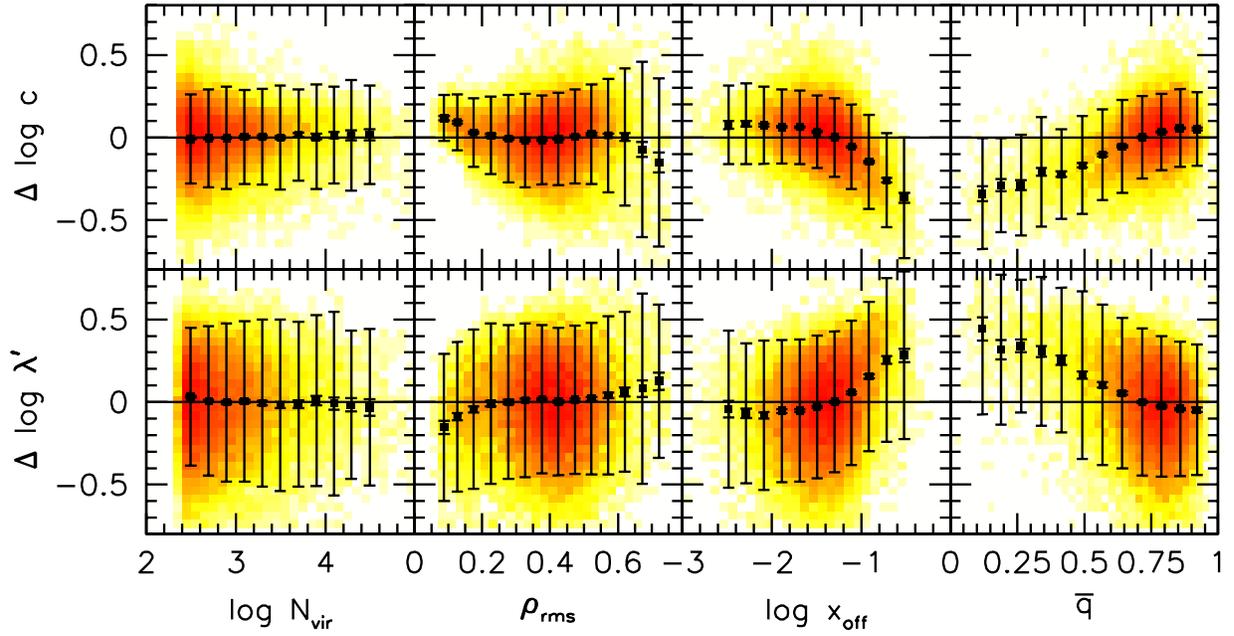,width=500pt}
\caption{Residuals of the $c-\Mvir$ (upper panels) and $\lamp-\Mvir$
  (lower  panels)  relations   as  fitted  in  Fig.~\ref{fig:clm_all},
  against $\Nvir$, $\rhorms$, $\xoff$, and $\qbar$ for all haloes
  with $\Nvir>250$. The large error  bars show twice the $1\sigma$ intrinsic
  scatter, while the small error bar shows the possion error on the mean.  }
\label{fig:dcl_all}
\end{figure*}

Fig.~\ref{fig:dcl_all}  shows  the  residuals  of  the  $c-\Mvir$  and
$\lamp-\Mvir$ relations for  haloes with $\Nvir>250$, against $\Nvir$,
$\rhorms$, $\xoff$,  and $\qbar$.  The  filled circles and  error bars
show the mean and 2$\sigma$ scatter  of points in equally spaced bins. 
The  smaller error bars,  sometimes barely  visible, show  the Poisson
error  on the  mean  ($\sigma/\sqrt{N}$).  Both  the  $c$ and  $\lamp$
residuals show  no trend with $\Nvir$,  even down to our  limit of 250
particles. This  indicates that numerical resolution  is not affecting
our results.

\begin{table*}
 \centering
 \begin{minipage}{140mm}
  \caption{Effect of RMS parameter on $c$ and $\lamp$ distributions for haloes with $\Nvir>10000$.}
  \begin{tabular}{lcccccccc}
\hline 
$\rhorms$ & N haloes & $<\log c_{12}>$ & $\sigma_{\ln c}$  & $<\log \lamp>$ & $\sigma_{\ln\lamp}$ & $<\rhorms>$ & $<\xoff>$ & $<\qbar>$ \\ 
\hline
 $>$ 0.25 & 88       & 0.770 & 0.392 & -1.285 & 0.528 & 0.36 & 0.100 & 0.52 \\
0.15-0.25 & 183      & 0.986 & 0.281 & -1.436 & 0.596 & 0.19 & 0.038 & 0.66 \\
0.10-0.15 & 226	     & 1.111 & 0.204 & -1.552 &	0.513 & 0.12 & 0.022 & 0.73 \\
0.00-0.10 & 117	     & 1.138 & 0.167 & -1.625 &	0.516 & 0.09 & 0.018 & 0.74 \\
\hline 
 $>$ 0.0  & 614	     & 1.039 & 0.353 & -1.493 &	0.597 & 0.17 & 0.031 & 0.68 \\
\hline

\end{tabular}
\end{minipage}
\end{table*}

Interestingly we  find that haloes  with the lowest $\rhorms$  tend to
have larger $c$, smaller scatter in $c$, and lower $\lamp$.  A similar
result for  halo $c$ was  found by Jing  (2000), who used  the maximum
deviation of the density profile from the NFW fit,
which is  similar to our $\rhorms$  parameter. In Table~2  we show the
effect of  the RMS parameter on the  mean and scatter of  halo $c$ and
$\lamp$.  Here  $<\log c_{12}>$  is the zero  point of the  $c- \Mvir$
relation, measured at $\Mvir = 10^{12}\hMsun$. To allow for comparison
with Jing  (2000) we only  use haloes with  $\Nvir > 10000$.   We find
that the highest  $\rhorms$ haloes have a mean $c$  which is roughly a
factor of 2 lower, and a scatter in $c$ which is roughly a factor of 2
higher,  than the  lowest  $\rhorms$ haloes,  in  agreement with  Jing
(2000).  Additionally  we also find  a factor of 2  difference between
the average $\lamp$ of the highest and lowest $\rhorms$ haloes. Haloes
with  the highest $\rhorms$  have the  highest mean  offset parameter,
$<\xoff>$, and  the lowest mean halo shape  parameter, $<\qbar>$. This
suggests that these haloes are the most unrelaxed.

The trends between $c$ and $\lamp$ with $\xoff$ are shown in the third
column  of Fig.~\ref{fig:dcl_all}.  These  show that  for haloes  with
small $\xoff$  the residuals are uncorrelated  with $\xoff$.  However,
these haloes  have concentrations that are higher  and spin parameters
that are lower than the overall average.  For $\xoff \gta 0.04$ ($\log
\xoff  > -1.4$) there  is a  clear systematic  trend that  haloes with
larger $\xoff$ have lower $c$ and higher $\lamp$.  The same trends are
seen for  $\qbar$, where  more prolate haloes  (of a given  mass) have
lower $c$ and higher $\lamp$.  This basically reflects the correlation
between $\xoff$ and $\qbar$ (see Fig.~\ref{fig:all5}).

These trends  are consistent with  unrelaxed haloes being  the systems
that experienced a recent major merger:  (i) the center of the halo is
poorly defined, which  results in a large $\xoff$  and an artificially
shallow (low  concentration) density profile,  (ii) the halo  shape is
more  prolate, and  (iii)  the spin  parameter  is higher  due to  the
orbital angular momentum `transferred' to the system due to the merger
(e.g.,  Vitvitska  \etal  2002;  Maller, Dekel  \&  Somerville  2002).
Ideally one would test the correspondence between merger histories and
$\xoff$   using   the  actual   merger   trees   extracted  from   the
simulation. Unfortunately,  we did not store sufficient  outputs to be
able to do  so.  We intend to address these issues  in a future paper,
based on a new set of simulations.

We now split our sample  into 4 sub-samples according to $\rhorms$ and
$\xoff$,  with  dividers  of  $\rhorms=0.4$  and  $\xoff=0.04$,  which
correspond to the mean of  the distributions of $\rhorms$ and $\xoff$,
respectively (see Fig.~\ref{fig:all5}).   We refer to the 4-subsamples
as:
\begin{itemize}
\item GOOD$\,$  ($\rhorms < 0.4$ \& $\xoff<0.04$);
\item BAD$\;\;\;\,$ ($\rhorms < 0.4$ \& $\xoff>0.04$);
\item UGLY$\,\,$  ($\rhorms > 0.4$ \& $\xoff>0.04$);
\item NOISY     ($\rhorms > 0.4$ \& $\xoff<0.04$).
\end{itemize}

\begin{figure*}
\psfig{figure=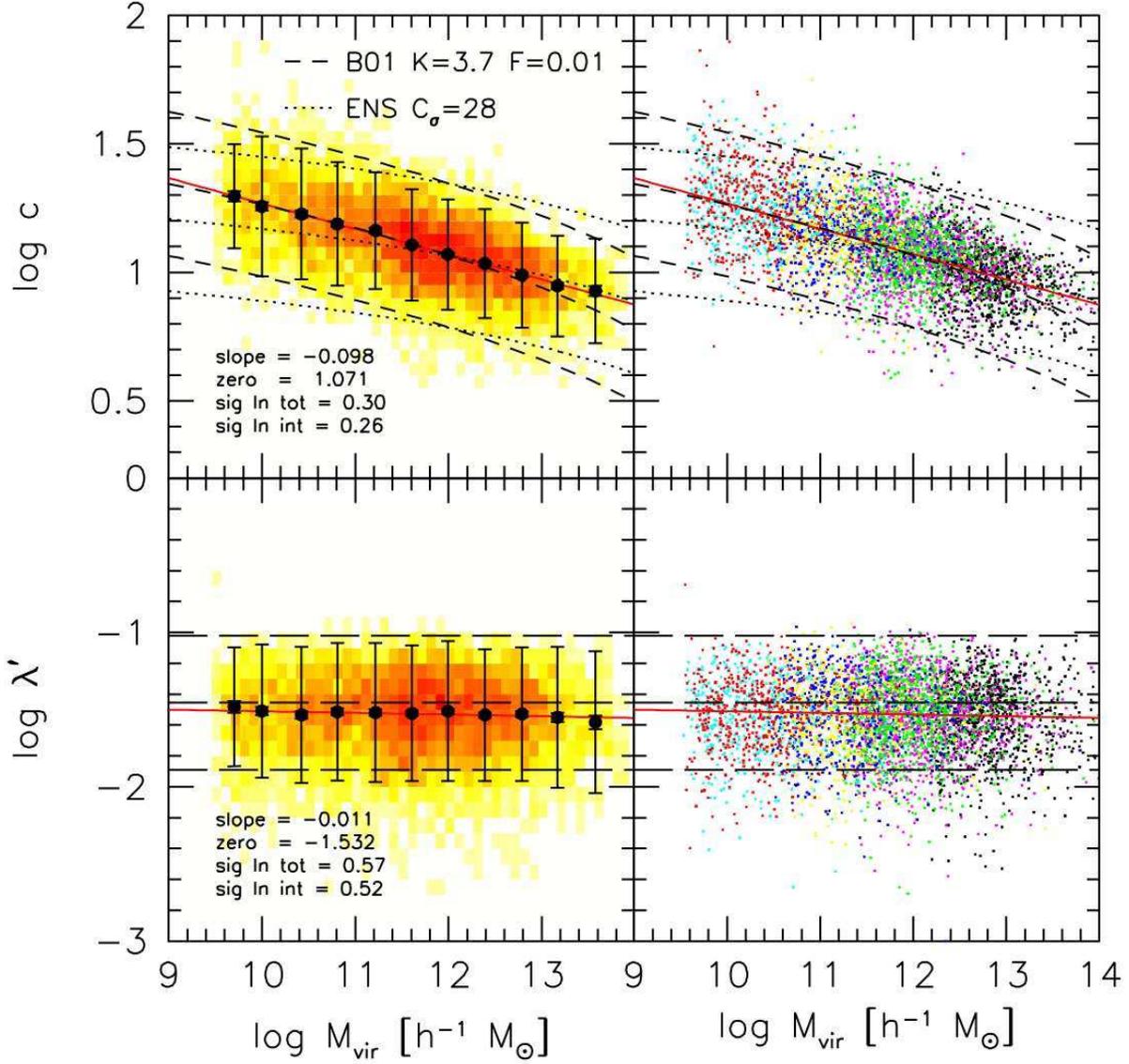,width=500pt}
\caption{As Fig.~\ref{fig:clm_all}, but for ``relaxed'' haloes ($\rhorms<0.4,
  \xoff<0.04$).      The     most     notable     differences     with
  Fig.~\ref{fig:clm_all} are  the reduced scatter  and shallower slope
  of the $c-\Mvir$ relation. See text for further details.}
\label{fig:clm_good}
\end{figure*}

Fig.~\ref{fig:clm_good}   shows   the   $c-\Mvir$  and   $\lamp-\Mvir$
relations for the haloes in  the GOOD sub-sample with $\Nvir>250$.  As
with the  full sample, this $c-\Mvir$  relation is well  fitted with a
single power-law given by
\begin{equation}
\label{goodc}
\log c = 1.071[\pm0.027] -0.098[\pm0.009](\log\Mvir -12).
\end{equation}
This relation  has a slope that  is $\sim 10$ percent  shallower and a
zero point that  is $\sim 10$ percent higher than  for the full sample
(eq.~[\ref{fullc}]).  The  total scatter  about this mean  relation is
$\sigma_{\ln c}=0.30\pm0.03$ and the intrinsic scatter is $\sigma_{\ln
  c}=0.26\pm0.03$, about $25$ percent lower  than for the full sample. 
The B01 model  again accurately fits the $c-\Mvir$  relation, but with
$K=3.7\pm 0.15$ (for $F=0.01$).   For $F=0.001$ the best-fit value for
$K$  is $2.9$.  

The  slope  of  the $c(M)$  relation  for  low  mass haloes  is  still
consistent with that  predicted by the B01 model  but steeper than the
prediction of  the ENS  model. Thus, the  fact that ENS  only compared
their model to  a small sample of well relaxed  haloes can not explain
the discrepancy  between their results  and those of B01.   As already
eluded  to above,  the  main reason  the  ENS model  was  found to  be
consistent  with their  own simulations  is that  the haloes  in their
simulation only covered  a small range in halo  masses, over which the
difference  in slope with  respect to  the B01  model is  difficult to
infer.

The slope of the $\lamp-\Mvir$ relation is again consistent with zero.
However, the median is $\sim 15$ percent lower ($\lamp=0.030\pm0.003$)
and   the  intrinsic   scatter  is   reduced  by   $\sim   5$  percent
($\sigma_{\ln\lamp}=0.52\pm0.01$).  These  differences  in  $c(\Mvir)$
between  the full set  of haloes  and our  GOOD sub-sample  are almost
identical  to those  obtained  by Wechsler  \etal  (2002) between  all
haloes and haloes without a major merger since $z=2$.  This reinforces
the notion  that our GOOD  sub-sample consists mostly of  haloes which
have not experienced a recent major merger.

\begin{figure*}
\psfig{figure=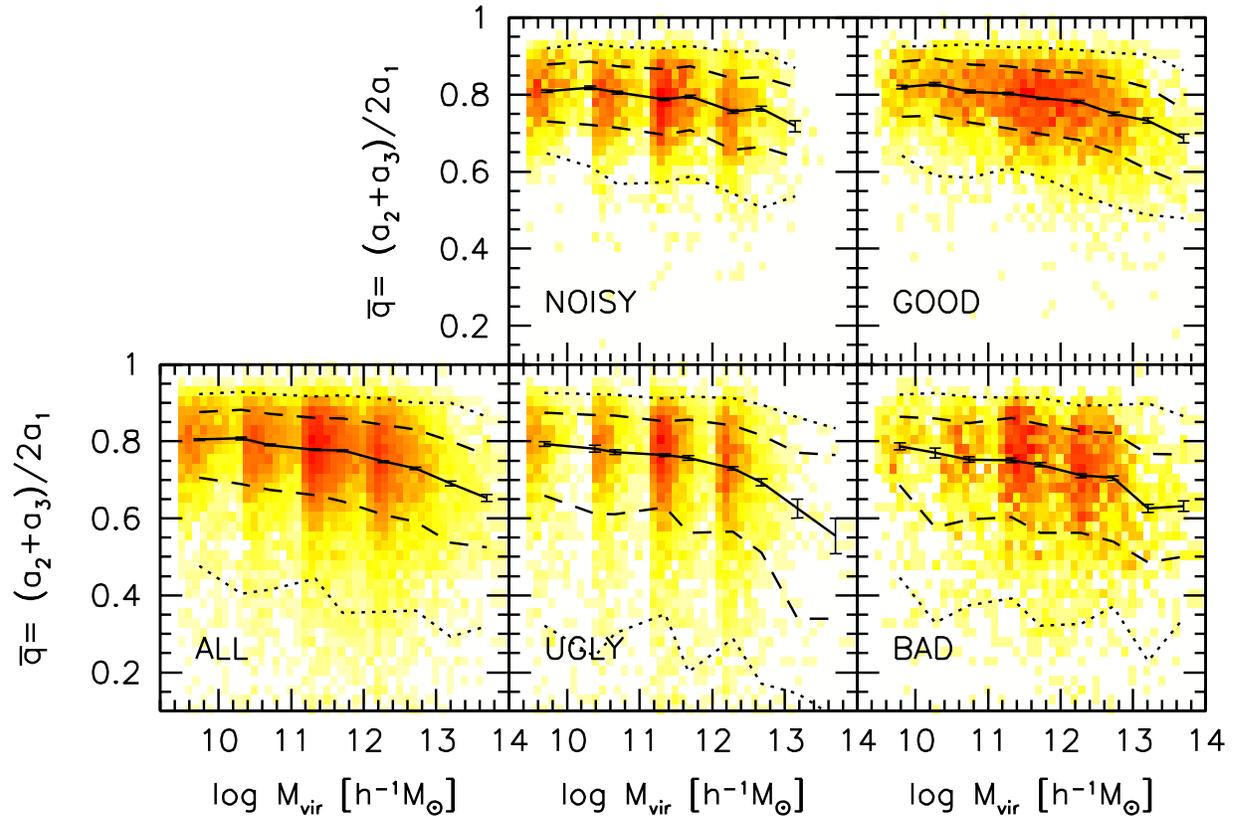,width=500pt}
\caption{Relation between $\qbar$ and $\Mvir$ for different
  sub-samples of haloes. The solid lines show the 50th percentile,
  dashed lines show the 16th and 84th percentiles, and the dotted
  lines show the 2.5th and 97.5th percentiles. The error bar gives the
  possion error on the median.}
\label{fig:qm5}
\end{figure*}

Fig.~\ref{fig:qm5} shows  the dependence  of the halo  shape parameter
$\qbar$ on halo mass.  We find that more massive haloes are on average
more flattened (more prolate),  in qualitative agreement with previous
studies (Jing \& Suto 2002;  Kasun \& Evrard 2005; Bailin \& Steinmetz
2005; Allgood \etal 2006).  As shown in Allgood \etal (2006), the halo
shape is fairly  tightly related to the halo  assembly time, such that
haloes that assemble later are  less spherical (and less relaxed).  To
first order this explains the decrease of $\qbar$ with increasing halo
mass,  as  well as  the  relation  between  $\qbar$ and  $\xoff$  (see
Fig.~\ref{fig:all5}). Fig.~\ref{fig:qm5}  also shows that  haloes with
high $\xoff$ (BAD  and UGLY sub-samples) have a  lower median $\qbar$,
as well as a much more pronounced tail to low $\qbar$.  Note also that
there are  very few highly  prolate haloes ($\qbar\lta 0.5$)  with low
$\xoff$  (NOISY  and   GOOD  sub-samples).   A  potentially  important
implication of this is that (LSB) disk galaxies, which are too fragile
to survive  in unrelaxed  haloes, are unlikely  to reside  in strongly
flattened haloes.


\section{Correlation between spin and concentration} 
\label{sec:clcorr}

In  their study of  the halo  angular momentum  distributions, Bullock
\etal (2001b) noticed that haloes with high $\lamp$ have concentration
parameters  that are  slightly lower  than average.   Although  such a
correlation  may  not  be  totally  unexpected, since  both  the  spin
parameter  and concentration  parameter depend  on the  mass accretion
history  (Wechsler \etal  2002; Vitvitska  \etal 2002),  Bullock \etal
argued that the $c$-$\lamp$ anti-correlation is a mere `artifact' from
the fact that  (i) $c$ and $\lambda$ are  uncorrelated (see also NFW),
and  (ii)   $\lambda$  and  $\lamp$   are  related  via   $f(c)$  (see
Section~\ref{sec:spinpar}).   However, Bailin  \etal  (2005) used  the
$\lampc$ definition for  the spin parameter (which should  be equal to
$\lambda$)  and claimed  a significant  anti-correlation  between halo
concentration  and  spin  parameter.   As discussed  in  Bailin  \etal
(2005), such  an anti-correlation may have  important implications for
the interpretation  of the rotation  curves of LSB disk  galaxies (see
discussion  in  Section~\ref{sec:intro}).  Using  our  large suite  of
simulations, we therefore re-investigate this issue.

\begin{figure*}
\psfig{figure=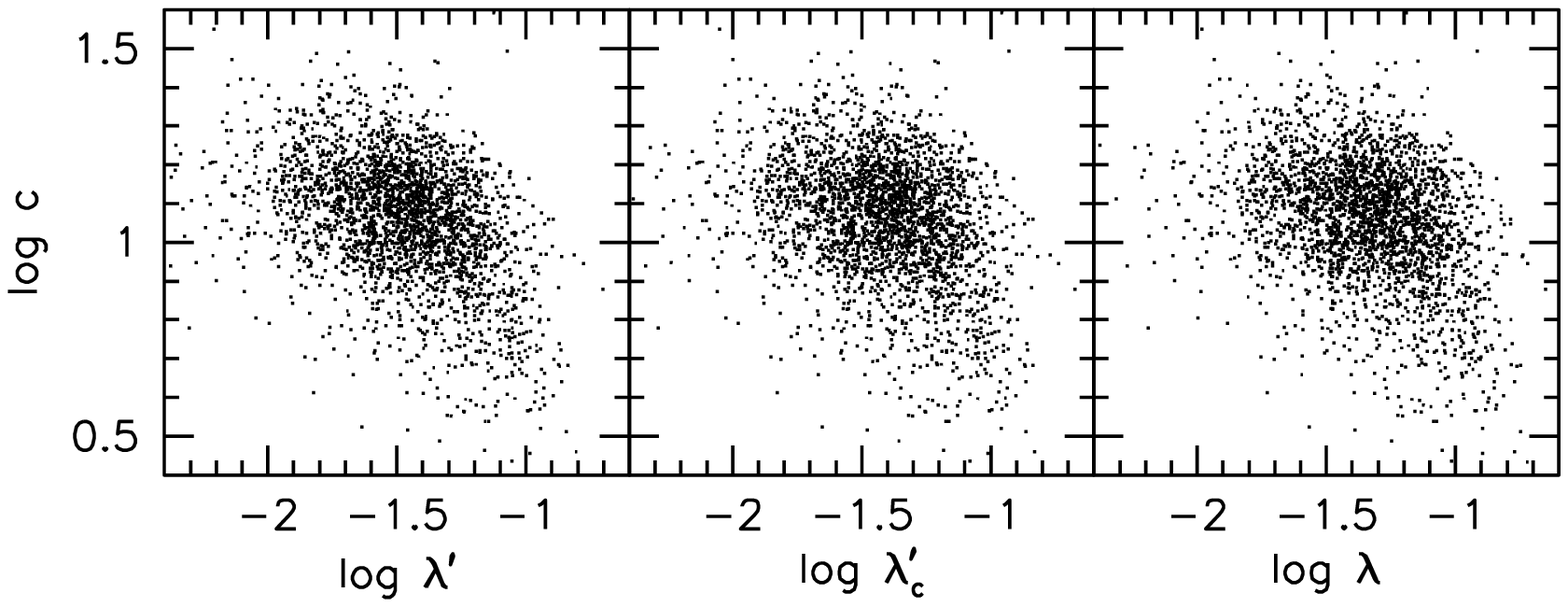,width=500pt}
\caption{Correlations between $c$ and $\lamp$, $\lampc$ and $\lam$ 
  for  haloes   with  $1\times  10^{11}  <  \Mvir   <  2\times  10^{12}
  \;h^{-1}\Msun $ and $N_{\rm vir} > 1000$. }
\label{fig:cl5}
\end{figure*}

Bailin \etal  (2005) focused on  haloes with $1\times  10^{11} h^{-1}
\Msun < \Mvir  < 2\times 10^{12} h^{-1}\Msun$ and  $N_{\rm FOF}>1290$. 
In Fig.~\ref{fig:cl5} we plot the  halo concentration as a function of
the  three  different  definitions  of the  spin  parameter;  $\lamp$,
$\lampc$ and $\lambda$. To facilitate the comparison with Bailin \etal
we only select haloes in  our simulations with $1\times 10^{11} h^{-1}
\Msun < \Mvir < 2\times  10^{12} h^{-1}\Msun$ and $\Nvir > 1000$.  All
three  plots reveal  a weak  but significant  correlation in  that the
lowest  concentration  haloes have  relatively  low  spin parameters.  
Contrary  to NFW  and Bullock  \etal  (2001b), but  in agreement  with
Bailin \etal  (2005), we  therefore argue that  $c$ and  $\lambda$ are
correlated. Note that resolution is not an issue here, since we obtain
the same result in each of our different simulations.

\begin{figure*}
\psfig{figure=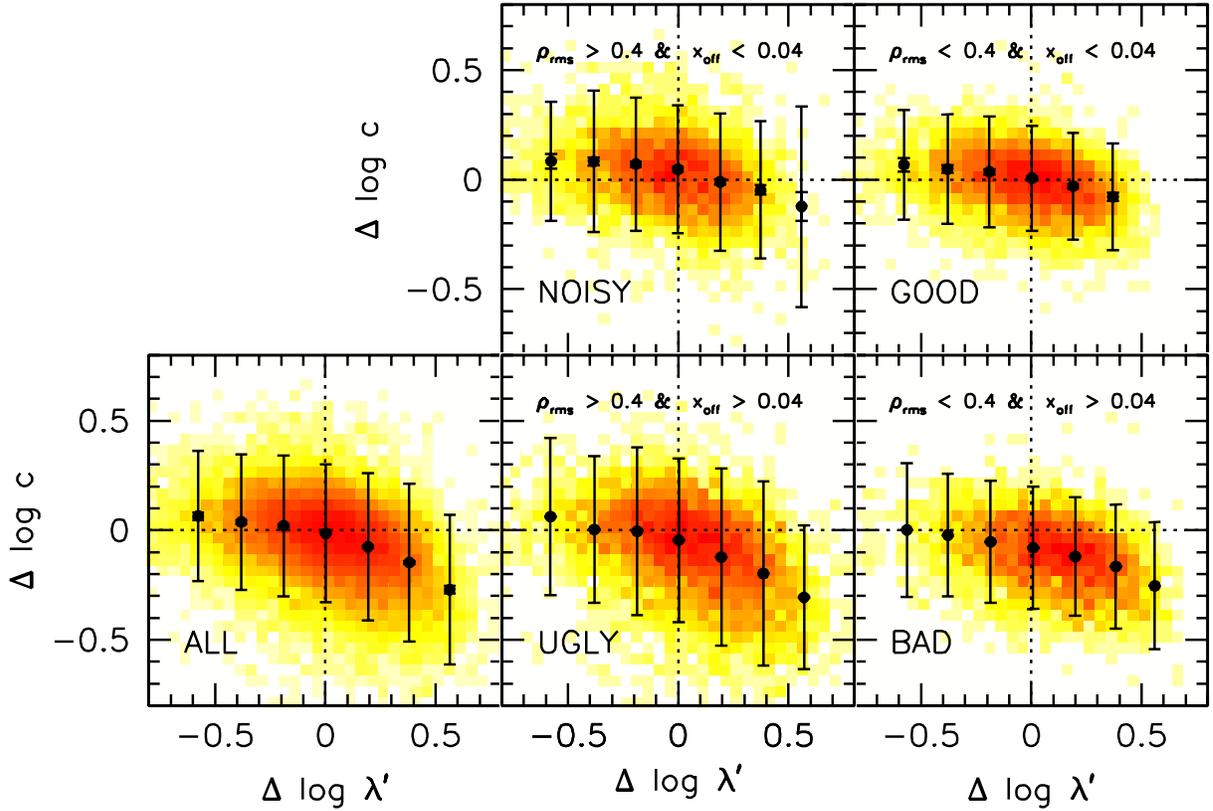,width=500pt}
\caption{Correlations between the residuals of the $c-\Mvir$ and 
  $\lamp-\Mvir$ relations  in Fig.~\ref{fig:clm_good}. The  lower left
  panel shows  all haloes,  while the remaining  panels show  the four
  sub-samples  defined according  to $\rho_{\rm  rms}$ and  $\xoff$ as
  indicated.  The  error bars  show the 2$\sigma$  scatter in  $c$ for
  each $\Delta\lamp$ bin.}
\label{fig:dcl5}
\end{figure*}

Due to the mass dependence of $c$, a more illustrative way to look for
a correlation between halo concentration and spin parameter is to plot
the residuals (at constant $\Mvir$) of the $c-\Mvir$ and $\lamp-\Mvir$
relations.  This  is shown in Fig.~\ref{fig:dcl5} for  all haloes with
$\Nvir>250$.  The  lower left panel  shows the residuals for  the full
sample, which reveals a clear trend that haloes with high $\lamp$ have
lower  than  average $c$.   We  now split  the  haloes  into the  four
sub-samples defined  above, and plot their residuals  (with respect to
the mean $c(M)$ and mean $\lamp$ for the GOOD sub-sample).  This shows
that the correlation  between $c$ and $\lamp$ is,  at least partially,
due to  the inclusion of haloes  with high $\xoff$  (i.e., haloes that
are  unrelaxed), independent of  whether the  halo has  a high  or low
$\rhorms$: the correlation is clearly more pronounced for the UGLY and
BAD sub-samples. However, the  GOOD and NOISY sub-samples still reveal
a small trend that haloes with larger $\lamp$ offsets have a lower $c$
offset.  We have repeated this analysis using $\lampc$ and $\lam$, and
the  small  correlation  between  residuals  in  the  GOOD  and  NOISY
sub-samples  remains.  We therefore  conclude that  there indeed  is a
small  intrinsic  correlation  between  $c$  and $\lamp$  at  a  fixed
$\Mvir$.  However, when excluding  the unrelaxed haloes, the amplitude
of  this correlation  is very  weak compared  to the  scatter  in both
parameters. In  Section~\ref{sec:sbdisk} below we  investigate to what
extent this small correlation may effect (LSB) disk galaxies.


\section{Environment dependence}

We now investigate whether the concentration, spin parameter and shape
of dark matter haloes are  correlated with the large scale environment
in which they are located.  This is interesting because disk galaxies,
and  LSBs  in  particular,  are  preferentially found  in  regions  of
intermediate to low density (Mo \etal 1994; Rosenbaum \& Bomans 2004).
If  haloes  in  low  density environments  have  different  structural
properties than  haloes of the  same mass in  over-dense environments,
this would therefore imply that  the haloes of (LSB) disk galaxies are
not a fair representation of the average halo population.

Using  a  set of  numerical  simulations,  for different  cosmologies,
LK99 studied  the environment dependence of  dark matter  haloes. 
The only  halo parameter  that was
found  to  be  correlated   with  environment  is  halo  mass  (denser
environments contain  more massive haloes).   Halo concentration, spin
parameter, shape and assembly redshift\footnote{The assembly time of a
  halo of mass $M_0$ is defined  as the time at which the most massive
  progenitor  has a  mass  equal to  $M_0/2$.}  were all  found to  be
uncorrelated with halo environment.

However, using exactly  the same simulations as LK99,  Sheth \& Tormen
(2004) found clear  evidence that  halo assembly times  correlate with
environment  based  on  a  marked correlation  function  analysis.   A
similar result was obtained by Gao, Springel \& White (2005), who used
the millennium  simulation (Springel  \etal 2005) to  demonstrate that
the low  mass haloes that  assemble early are more  strongly clustered
than haloes of the same mass that assemble later.  For massive haloes,
however, the  dependence on the assembly  time was found  to diminish. 
This has  since been  confirmed by a  number of studies  (Harker \etal
2006; Zhu  \etal 2006; Wetzel \etal  2007; Jing, Suto \&  Mo 2007). In
addition, it has been found that the clustering of haloes also depends
on  other  halo properties,  such  as  halo  concentration, halo  spin
parameter,  subhalo properties,  and  the time  since  the last  major
merger (Wechsler \etal 2006; Zhu \etal 2006; Wetzel \etal 2007; Gao \&
White  2007; but  see  also Percival  \etal  2003), which  is not  too
surprising  given  that  all  these  properties  correlate  with  halo
formation time  (e.g., Wechsler \etal  2002; van den Bosch,  Tormen \&
Giocoli 2005). All these results seem to overrule the finding by LK99,
and  seem  to  suggest  that  dark  matter  properties  depend  rather
sensitively on their large scale environment.

\begin{figure*}
\psfig{figure=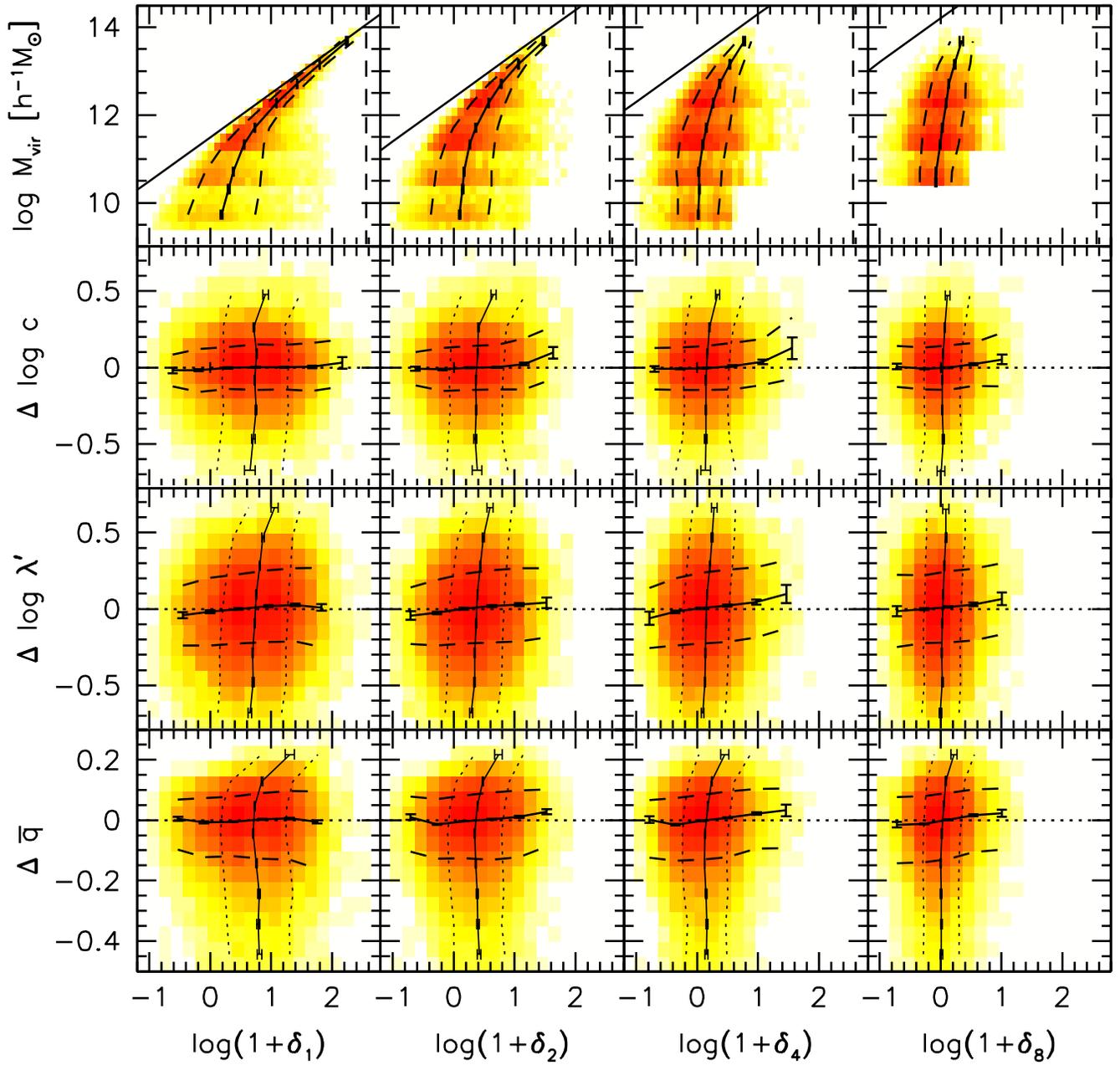,width=500pt}
\caption{Environment dependence of $\Mvir$, and of the residuals, at
  fixed  $\Mvir$  of the  $c-\Mvir$,  $\lamp-\Mvir$ and  $\qbar-\Mvir$
  relations.  Here  environment is  defined by the  matter overdensity
  within spheres of radii $1,2,4$  and $8 \;h^{-1}Mpc$. In the top row
  the  diagonal line shows  the minimum  overdensity for  a halo  of a
  given mass,  and the vertical  dashed line shows the  overdensity at
  the virial  radius. In each panel,  the solid curve  shows the mean,
  the  error bars indicate  the Poisson  errors on  the mean,  and the
  dashed curves show the $\pm 1\sigma$ scatter in each mass bin.
  In the second, third and fourth rows the almost vertical lines 
  indicate the  average overdensity (plus  its  error)  for haloes in a given 
  residual bin.}

\label{fig:mclomega_all}
\end{figure*}

Using our  large sample of objects, we investigate  whether the
concentration, spin parameter and shape of haloes are correlated with
the large  scale environment in  which they are located.   Rather than
using (marked) correlation functions, we follow LK99 and correlate the
halo  properties  with  the  overdensity,  $1+\delta_r$,  measured  in
spheres of radius $r$(with $r=1,2,4,8\, \hMpc$) centered on each halo:
\begin{equation}
1+\delta_r = \frac{<\rho(r)>}{\rho_{\rm u}}=\frac{3 M(<r)}{4 \pi r3}
\frac{1}{\rho_{\rm crit} \OmegaM }.
\end{equation}
The results  for all haloes are shown  in Fig.~\ref{fig:mclomega_all}. 
The top  row shows the relation  between virial mass  and overdensity. 
The vertical dashed  line shows the overdensity at  the virial radius. 
The solid diagonal line indicates the  mass scale $M = {4 \over 3} \pi
(1 +  \delta_r) \rho_{\rm  crit} \OmegaM r^3$.   Thus all  haloes with
$\Rvir$ less than the filter radius  should lie below this line, as is
the case.  Note that we do  not include sub-haloes in our analysis, so
the only haloes that  can have very high densities $\log(1+\delta)\gta
2$ must be haloes with virial  radii close to the filter radius, as is
the  case.  We  see that  more  massive haloes  tend to  live in  more
overdense regions (cf.  LK99).  As  the filter radius is increased the
mean overdensity tends towards zero, and the scatter in overdensity is
reduced, as expected.  Note that  we do not compute the overdensity on
$8 h^{-1}  \rm Mpc$  scales in the  simulations with the  smallest box
size.

The second, third and  fourth rows of Fig.~\ref{fig:mclomega_all} show
the residuals  (at fixed $\Mvir$) of  the $c$-$\Mvir$, $\lamp$-$\Mvir$
and  $\qbar$-$\Mvir$  relations,  respectively,  all  as  function  of
overdensity. The roughly horizontal  lines with errorbars indicate the
mean residual (plus its errors)  as function of overdensity, while the
dashed lines outline the $\pm  1\sigma$ scatter. Note that, within the
errors  on the  mean,  there  is no  significant  indication that  the
residuals  are  larger  or  smaller  in  regions  that  are  over-  or
under-dense. In other words,  these results are in excellent agreement
with  those of  LK99  and seem  to  suggest that  halo properties  are
independent of  their large scale environment.  Note  that our results
cover  a much  wider dynamic  range  in halo  mass, and  are based  on
higher-resolution simulations than in the  case of LK99.  How can this
be  reconciled with  the findings  based on  the  correlation function
analyses  described above?  Some  insight is  provided by  the roughly
vertical lines with errorbars (in the second, third and fourth rows of
Fig.~\ref{fig:mclomega_all}),  which indicate the  average overdensity
(plus  its  error)  for haloes in a given residual bin. The
corresponding dotted  lines outline  the $\pm 1\sigma$  scatter. These
show  that  haloes  with   the  largest  concentration,  largest  spin
parameter, and/or  that are  most spherical (all  with respect  to the
average for their  mass) are located in slightly  denser environments. 
Since the correlation function  reflects ensemble averages, this seems
consistent  with the  findings  that more  concentrated haloes  and/or
those with  a larger  spin parameter are  more strongly  clustered. We
emphasize, though, that the trends seen in Fig.~\ref{fig:mclomega_all}
are (i) weaker when environment is measured over a larger volume, (ii)
only reveal an environment dependence  at the extremes of the residual
distributions, and (iii)  largely vanish when we only  focus on haloes
in our GOOD sub-sample (not shown).

\begin{figure*}
\psfig{figure=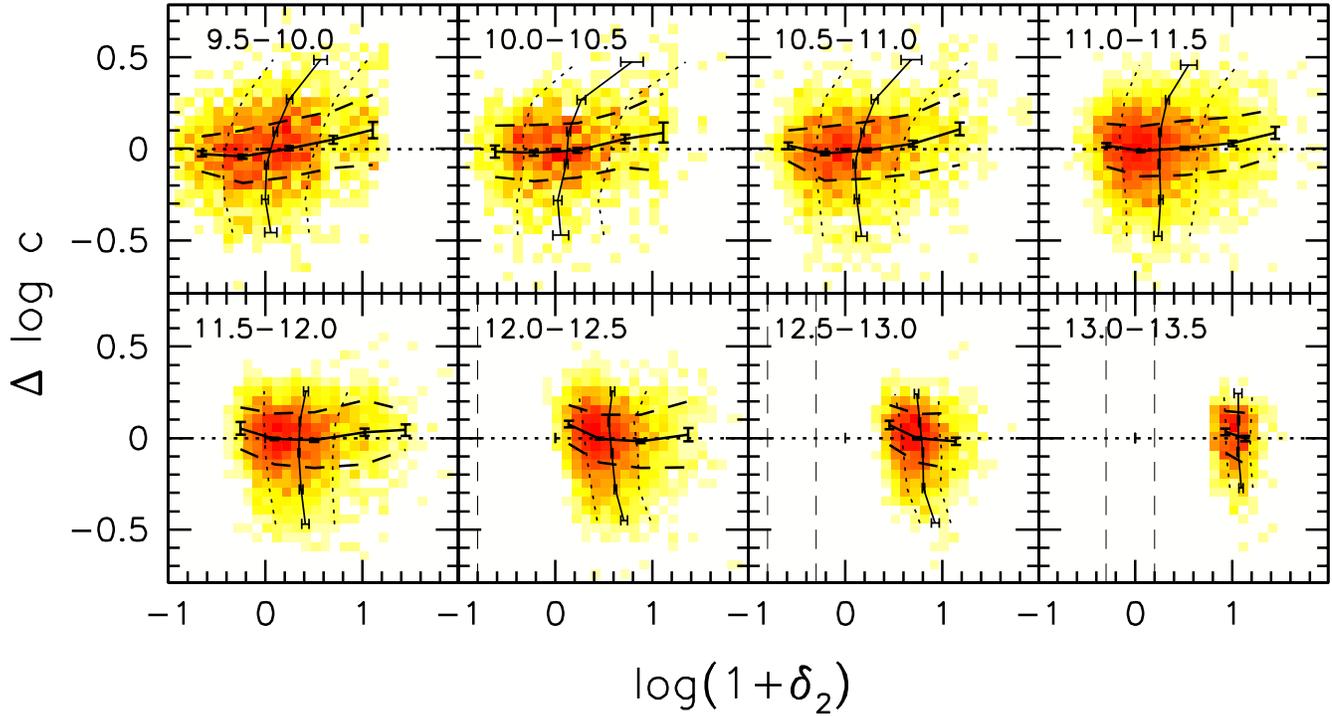,width=500pt}
\caption{Environment dependence of the residuals, at fixed $\Mvir$, 
  of the $c-\Mvir$ relation.  Different panels correspond to different
  halo     mass    bins,     as     indicated    (numbers     indicate
  $\log[\Mvir/h^{-1}\Msun]$). The solid curve shows the mean, the error
  bars show the Poisson error on  the mean, and the dashed curves show
  the  $\pm 1\sigma$  scatter.   The dashed  lines show  the
  overdensity corresponding to a single  halo in a sphere of radius $2
  \hMpc$ with a mass equal to  the lower and upper mass limits of each
  panel. The almost vertical lines have the same meaning of Fig.~\ref{fig:mclomega_all}
  indicating the  average overdensity (plus  its  error)  for haloes in a given 
  residual bin.}
\label{fig:dclomega}
\end{figure*}

As  emphasized  in  Harker  \etal  (2006),  in  order  to  reveal  the
correlation between environment and  assembly time, it is important to
only consider haloes in a  relatively small mass range.  Therefore, in
Fig.~\ref{fig:dclomega}  we  plot   the  residuals  of  the  $c-\Mvir$
relation versus  overdensity in $2  \hMpc$ spheres for haloes  in mass
bins with  a width of 0.5  dex.  The various lines  and errorbars have
the same  meaning as in Fig.~\ref{fig:mclomega_all}.   For haloes with
$\Mvir \lta 10^{12} h^{-1} \Msun$  there is a weak, mildly significant
indication  that haloes  in denser  environments have  slightly higher
concentrations   (reflected  by   close-to-horizontal   lines).   More
pronounced, however,  is the trend that more  concentrated haloes live
on  average  in denser  environments  (reflected by  close-to-vertical
lines). This is  consistent with the fact that  more concentrated, low
mass haloes are more strongly clustered (cf. Wechsler \etal 2006).  It
is clear from  Fig.~\ref{fig:dclomega} that the environment dependence
is weaker  for more  massive haloes;  in fact for  haloes in  the mass
range  $12  \lta  \log  [M/(h^{-1}\Msun)] \lta  13.5$  no  significant
environment dependence is seen (this explains why the signal is weaker
in Fig.~\ref{fig:mclomega_all} where we  add all the masses together).
Again  this  is in  good  agreement  with  the clustering  results  of
Wechsler  \etal (2006) and  Gao \&  White (2007),  who found  that the
dependence on halo concentration vanishes for haloes with masses close
to the  typical collapsing mass,  $M^*$: for the cosmology  assumed in
our simulations  $M^* = 6.7  \times 10^{12} h^{-1} \Msun$.   Thus, our
results  are  in good  agreement  with  the  various claims  based  on
correlation function analyses.  They also illustrate, though, that the
trends are weak  compared to the scatter, which  explains why LK99 did
not  notice  any  environment  dependence.  Only  when  one  carefully
estimates the average environment  as function of halo property, which
is  what  a  correlation   function  measures,  does  the  environment
dependence reveal itself.

\begin{figure*}
\psfig{figure=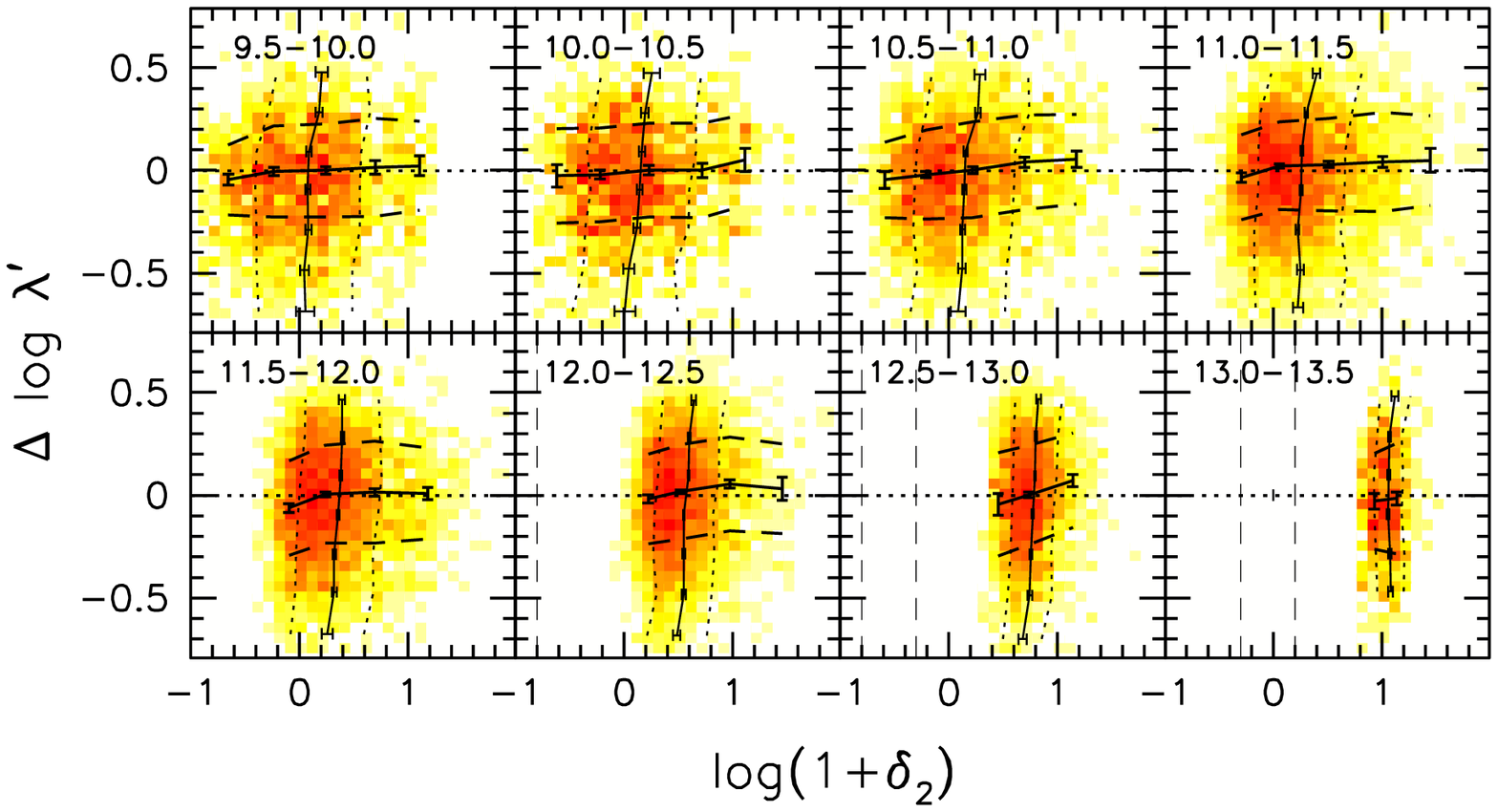,width=500pt}
\caption{Same as Fig.~\ref{fig:dclomega}, except that here we show the
  residuals of the $\lamp-\Mvir$ relation.} 
\label{fig:dclomegab}
\end{figure*}

Figure~\ref{fig:dclomegab} shows  the residuals, at  fixed $\Mvir$, of
the $\lamp-\Mvir$  relation as function of  $\log(1+\delta_2)$ for the
same   mass   bins   as   in  Fig.~\ref{fig:dclomega}.    Unlike   the
concentration, the  spin parameter reveals  no significant environment
dependence, in  any mass bin. This  is inconsistent with  Gao \& White
(2007) who found  that higher spin haloes are  more strongly clustered
than low spin haloes of the  same mass. It is unclear why our analysis
of the environmental  densities does not recover this  trend, while it
does reproduce the trend for the halo concentrations.

\begin{figure*}
\psfig{figure=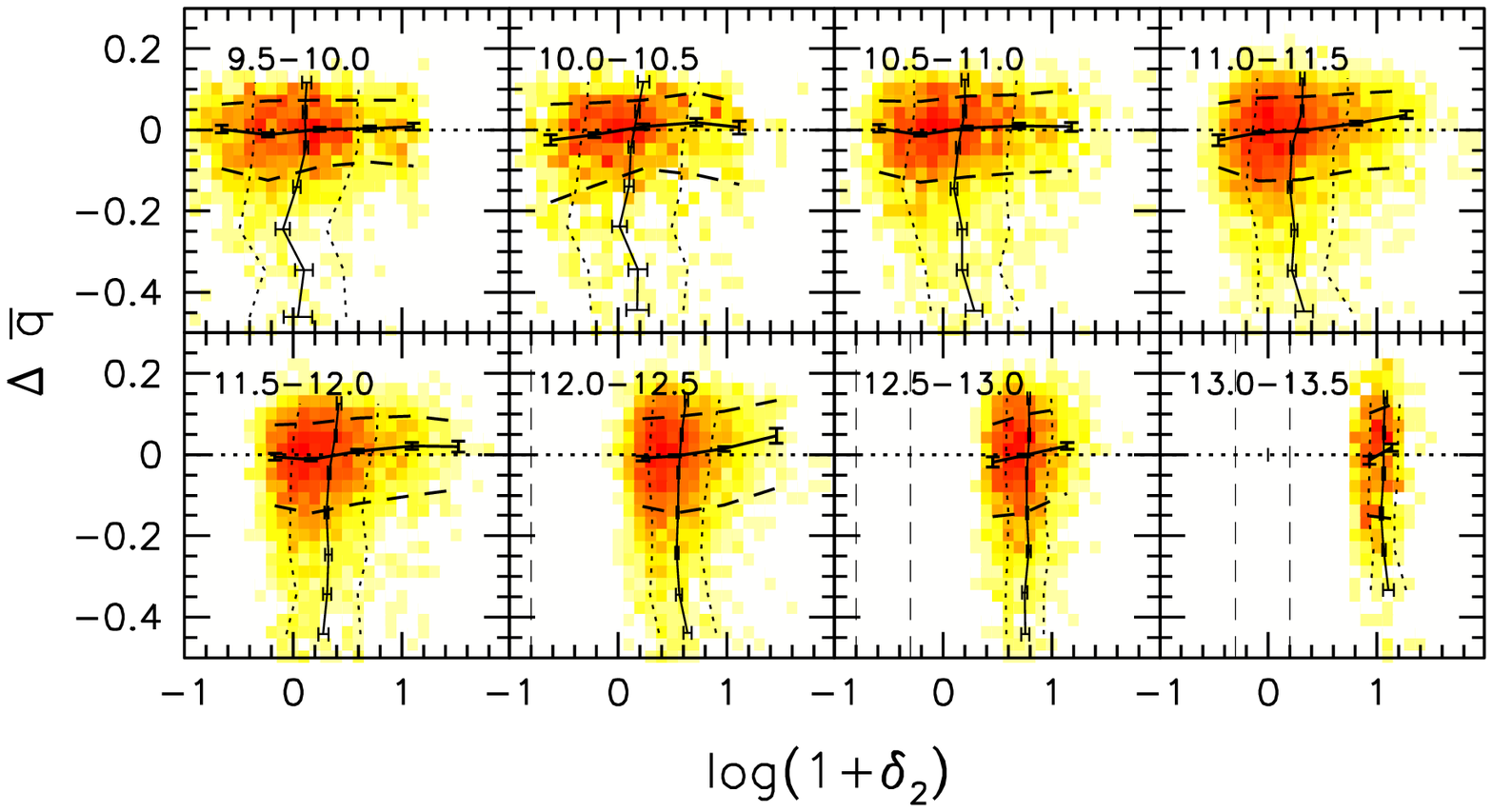,width=500pt}
\caption{Same as Fig.~\ref{fig:dclomega}, except that here we show the
  residuals of the $\qbar-\Mvir$ relation.} 
\label{fig:dclomegac}
\end{figure*}

Finally,     Fig.~\ref{fig:dclomegac}     shows     the    same     as
Figs~\ref{fig:dclomega}  and~\ref{fig:dclomegab}  but  for  the  shape
parameter $\qbar$.  As for the  spin parameter, there is no indication
for any  significant environment depencence.  This seems  at odds with
Fig.~\ref{fig:mclomega_all}  (fourth row), which  shows that  the most
spherical haloes  (those with large  positive $\Delta \qbar$)  live in
denser environments.  This apparent discrepancy simply reflects number
statistics: only the haloes in the  bin with $\Delta \qbar > 0.2$ seem
to reside in regions that are denser than average.  When split in mass
bins, however, there  are too few haloes with $\Delta  \qbar > 0.2$ to
reveal the signal.


\section{The Host Haloes of LSB Disk Galaxies}
\label{sec:sbdisk}

We now investigate whether LSB disk galaxies are expected to reside in
dark  matter haloes  that  form a  biased  sub-set in  terms of  their
concentration parameters. In the MMW model the central surface density
of  an   exponential  disk,  $\Sigma_{0,\rm  d}$,   is  determined  by
$\lambda$, $c$, $\Mvir$, and the galaxy mass fraction $\mgal$ (defined
as the ratio between disk  mass and halo mass). A lower $\Sigma_{0,\rm
  d}$ will result from
\begin{itemize}
\item a higher $\lambda$ at fixed  $c$, $\Mvir$, and $\mgal$;
\item a lower $\Mvir$ at fixed $c$, $\lambda$, and $\mgal$;
\item a lower $c$ at fixed $\lambda$, $\Mvir$, and $\mgal$;
\item a lower $\mgal$ at fixed $c$, $\lambda$, and $\Mvir$.
\end{itemize}
To complicate  matters, lower mass haloes  have higher concentrations,
$\lambda$     and    $c$     are    weakly     anti-correlated    (see
Section~\ref{sec:clcorr}),  and $\mgal$ is  expected to  increase with
$\Mvir$ due to various astrophysical feedback processes.

To  investigate the  impact  of  all these  relations  on the  surface
brightness of disk galaxies, we construct MMW type models as described
in Dutton \etal (2007).  These  models consist of an exponential disk,
a Hernquist  bulge and a NFW  halo.  The halo  parameters are $\Mvir$,
$c$ and  $\lampc$ ($\equiv\lambda$), which  we take directly  from the
haloes of our  GOOD sub-sample. An additional parameter  is the galaxy
mass fraction  $\mgal$, which we  fix to $\mgal=0.04$ for  simplicity. 
The bulge formation recipe is based on disk instability, and therefore
only effects the highest surface brightness disks; the details of this
bulge formation  recipe are  not important for  this work.   We assume
that the halo is unaffected by  the formation of the disk, i.e.  we do
not consider  adiabatic contraction.   As highlighted in  Dutton \etal
(2007), models with adiabatic contraction are unable to simultaneously
match  the zero-point  of  the Tully-Fisher  relation  and the  galaxy
luminosity function.

The MMW formalism gives the  galaxy mass, $M_{\rm gal}$, baryonic disk
scale length, $R_{\rm d}$, and central surface density of the baryonic
disk, $\Sigma_{0,  \rm d}$.  As  described in Dutton \etal  (2007), we
split the  disk into a stellar  and a gaseous  component assuming that
all  disk material  with a  surface density  $\Sigma(R)  > \Sigma_{\rm
  crit}(R)$ has  been turned into  stars. Here $\Sigma_{\rm  crit}$ is
the star formation threshold  density, modeled as the critical surface
density given by Toomre's  stability criterion (Toomre 1964; Kennicutt
1989).The resulting stellar surface  density profile is fitted with an
exponential profile  to obtain the  scale length of the  stellar disk,
$R_*$,  and   the  central  surface  density  of   the  stellar  disk,
$\Sigma_{0,*}$.  Note that in general $R_*< R_{\rm d}$.

In  order to  facilitate  a comparison  with  observations we  convert
$\Sigma_{0,  \rm  *}$  into   $\mu_{0,\rm  B}$,  the  central  surface
brightness of the stellar disk in the B-band, using the B-band stellar
mass-to-light ratio, $\Upsilon_{\rm B}$.  Using data from Dutton \etal
(2007) and relations between mass-to-light  ratios and color from Bell
\etal (2003) we obtain
\begin{equation}
\log \Upsilon_B = 0.06 + 0.25 \log\left( {M_* \over 10^{10} \Msun}
  \right) 
\end{equation}
where we have  adopted a Kennicutt IMF and a  Hubble constant $h=0.7$. 
In principle this relation has a scatter of $\simeq 0.1$ dex, which we
ignore for simplicity.

\begin{figure*}
\psfig{figure=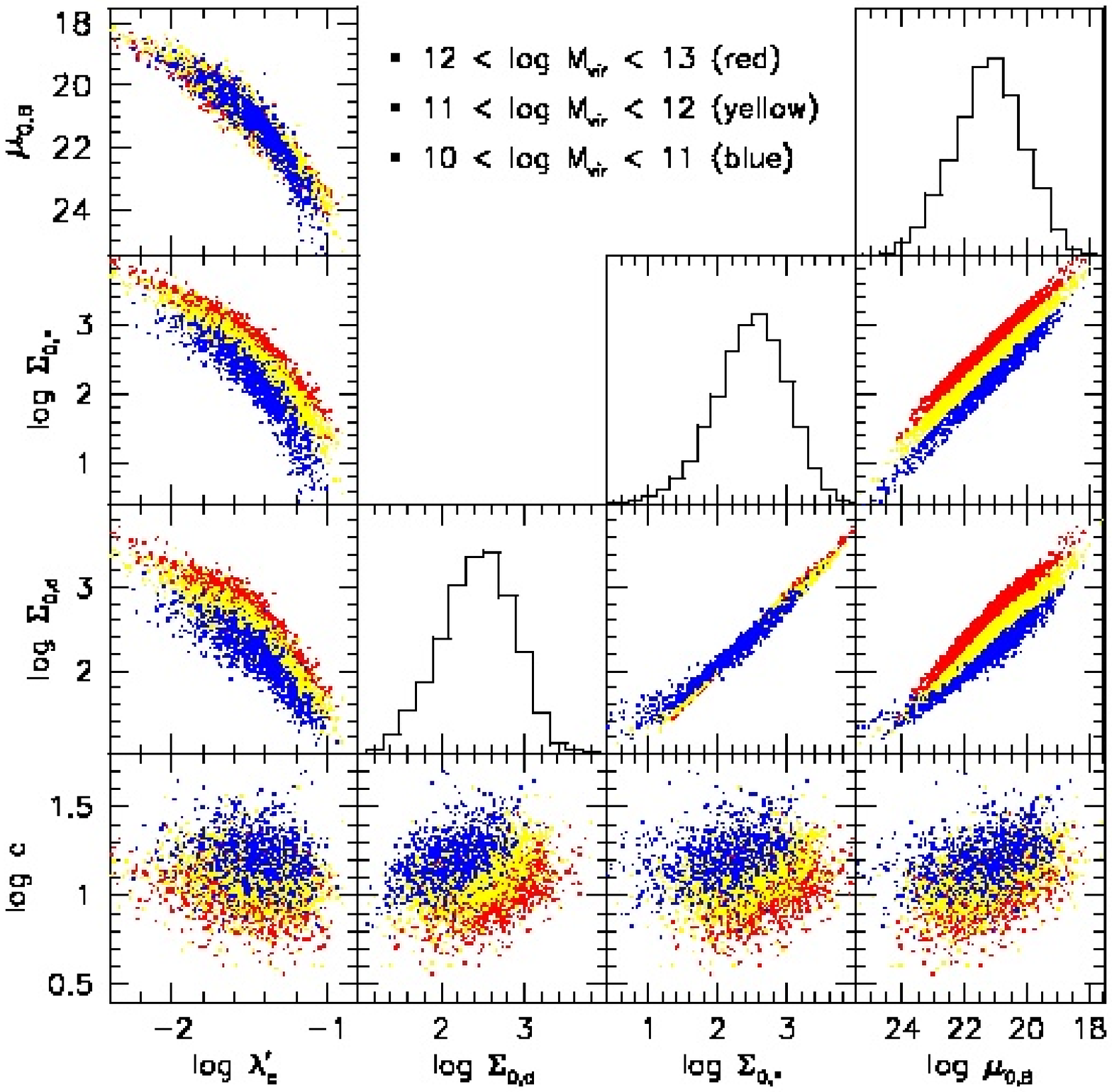,width=500pt}
\caption{Correlations between the halo variables $\lampc$ and $c$
  with  the  galaxy  properties  $\Sigma_{0}$ (baryonic  disk  central
  surface  density, units  of $\Msun  \,\rm  pc^{-2}$), $\Sigma_{0,*}$
  (stellar disk central  surface density, units $\Msun\,\rm pc^{-2}$),
  and  $\mu_{0,B}$ (stellar  disk  central surface  brightness in  the
  B-band,   units   $\rm   magn.\,arcsec^{-2}$),   for   haloes   with
  $\Nvir>250$, $\rhorms<0.4$, and $\xoff<0.04$ (i.e.  relaxed haloes).
  The points are color coded  according to $\Mvir$ (units $\hMsun$) as
  indicated.}
\label{fig:scl2-good}
\end{figure*}

Fig.~\ref{fig:scl2-good}  shows  correlations  between $c$,  $\lampc$,
$\Sigma_{0,\rm  d}$  (central   surface  density  of  baryonic  disk),
$\Sigma_{0,*}$  (central   surface  density  of   stellar  disk),  and
$\mu_{0,B}$ (central surface brightness  in the B-band) and histograms
of $\Sigma_{0,\rm d}$, $\Sigma_{0,*}$  and $\mu_{0,B}$ for GOOD haloes
in  the mass  range $10^{10}  h^{-1} \Msun  < \Mvir  <  10^{13} h^{-1}
\Msun$.  The  distributions of $\Sigma_{0,\rm  d}$, $\Sigma_{0,*}$ and
$\mu_{0,\rm   B}$  are   approximately   log-normal,  reflecting   the
log-normal distribution of  $\lampc$.  The distribution of $\mu_{0,\rm
  B}$ has  a peak value in  agreement with the Freeman  value of 21.65
magn. arcsec$^{-2}$.

As expected, there is  a strong correlation between $\Sigma_{0,\rm d}$
and $\lampc$,  in that haloes  with larger spin parameters  host lower
surface density disks. The flattening of this relation at low $\lampc$
owes to  our bulge formation  recipe. At fixed $\lampc$,  more massive
haloes  have higher  $\Sigma_{0,\rm  d}$, despite  their (on  average)
smaller concentrations.  The scatter in $c$ at a fixed $\Mvir$ results
in some  overlap between  the three mass  samples, but the  three mass
ranges are  clearly visible.  Thus, the dependence  of surface density
on  halo mass  is at  least  as important  as the  dependence on  halo
concentration.   The same  trends  are seen  in  the relation  between
surface density of the {\it stellar} disk and spin parameter.  However
the  mass separation  is no  longer  present in  the relation  between
surface brightness  and spin parameter.   This is because, at  a given
$\Sigma_{0,*}$, more massive  haloes have higher stellar mass-to-light
ratios, and hence lower surface brightness.

The  lower  left  panel  of Fig.~\ref{fig:scl2-good}  plots  the  halo
concentration  versus   the  halo  spin   parameter.   Although  these
parameters seem uncorrelated, there is a weak anti-correlation between
$c$ and  $\lampc$, as discussed in  Section~\ref{sec:clcorr}. The fact
that this  correlation is not as pronounced  as in Fig.~\ref{fig:cl5},
is due  to the fact  that here we  only consider the GOOD  sub-sample. 
The other three  panels in the lower row  show the correlation between
halo  concentration and  disk  surface density  (or brightness).   For
haloes of  a given  mass, there  is a clear  correlation in  that more
concentrated haloes  host higher  surface density (brightness)  disks. 
This  correlation has  two  origins: centrifugal  equilibrium and  the
(weak)  correlation between  $\lampc$  and $c$.   In  what follows  we
investigate the relative importance of both of these causes.

\begin{figure*}
\psfig{figure=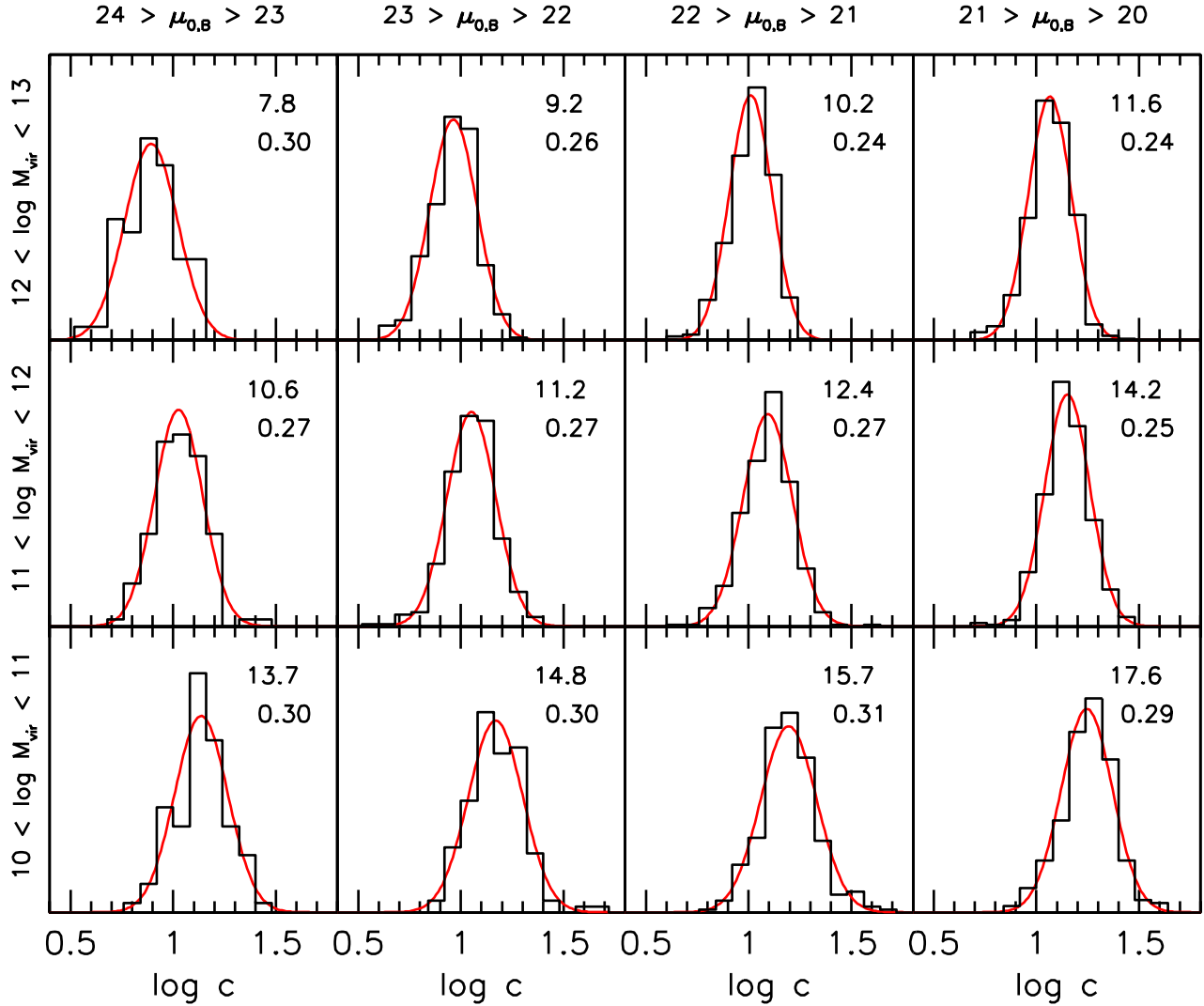,width=500pt}
\caption{Histograms of concentration parameter for different viral
  masses  (units of  $\hMsun$) and  B-band central  surface brightness
  (units  of  $\rm  magn.\,arcsec^{-2}$)  for the  model  galaxies  in
  Fig.~\ref{fig:scl2-good}.   The  three  mass  ranges  (from  top  to
  bottom)  correspond  to massive,  intermediate  and dwarf  galaxies,
  while  the  four surface  brightness  ranges  (from  left to  right)
  correspond  to  low,  intermediate,   high  and  very  high  surface
  brightness. For each subset of  haloes the red line shows a gaussian
  distribution with a mean $c$ and standard deviation $\sigma_{\ln c}$
  as  given in  the top  right corner  of each  panel.  Note  that the
  dependence on mass  is much stronger than the  dependence on surface
  brightness.}
\label{fig:scl3-good}
\end{figure*}

Fig.~\ref{fig:scl3-good} shows  the distribution of  $c$ for different
ranges of $\Mvir$ and $\mu_{0,\rm  B}$.  The three mass ranges roughly
correspond  to massive  galaxies  ($150\lta \Vvir  \lta  300 \kms  $),
intermediate mass galaxies ($70 \lta  \Vvir \lta 150 \kms$), and dwarf
galaxies  ($30\lta   \Vvir  \lta  70  \kms$).   At   a  fixed  surface
brightness, less  massive haloes  have higher $c$  as expected.   At a
fixed halo mass, there is  a clear trend that lower surface brightness
disks reside in  less concentrated haloes.  If we  define LSB galaxies
as those with a central surface brightness $24 \gta \mu_{0,\rm B} \gta
23 \,{\rm magn.} {\rm arcsec}^{-2}$, we find that they live in a sub-set
of haloes  whose average concentration  (at fixed halo mass)  is $\sim
15$ percent  lower than  the overall average  for that halo  mass.  We
therefore  conclude  halo concentrations  inferred  from LSB  rotation
curves should not be compared  to $\langle c \rangle_M$, but rather to
$f \langle c\rangle_M$, with $f \simeq 0.85$ a bias correction factor.
A similar conclusion was obtained  by Bailin \etal (2005), except that
they found a bias correction  factor of $f \simeq 0.70$.  As discussed
in Section~\ref{sec:clcorr}  this owes to  the fact that they  did not
remove  unrelaxed haloes  from  their sample.   Since  we consider  it
unlikely that LSB  galaxies reside in unrelaxed haloes,  we belief our
bias correction factor to be more realistic.

Finally, in order to investigate the  origin of this bias, we have run
a  control sample  with the  same  distributions of  $\Mvir$, $c$  and
$\lampc$ as the  simulation data, but with no  correlation between $c$
and $\lampc$.  This reduces the  correction factor to $f \simeq 0.95$,
and  therefore shows that  the main  contribution to  $f$ owes  to the
(very weak) anti-correlation between  halo concentration and halo spin
parameter.    The   remaining   contribution  simply   reflects   that
centrifugally supported disks in less concentrated haloes will be less
concentrated themselves.


\section{Summary}
\label{sec:conclusion}

In this paper we have used a set of cosmological N-body simulations to
study   the  scaling   relations,  at   redshift  zero,   between  the
concentration   parameter,  $c$,   spin   parameter,  $\lamp$,   shape
parameter,  $\qbar$, and  mass, $\Mvir$,  of  a large  sample of  dark
matter haloes.  Due  to the combined set of  simulations, we were able
to  extend previous  studies to  a  mass range  at least  an order  of
magnitude smaller, covering the full range of masses in which galaxies
are expected to  form: $3\times 10^{9} \hMsun \lta  \Mvir \lta 3\times
10^{13} \hMsun$.

For  this mass  range  we find  $c\propto\Mvir^{-0.11}$,  which is  in
agreement with the model of Bullock \etal (2001a), but in disagreement
with the  model of Eke, Navarro  \& Steinmetz (2001)  which predicts a
significantly shallower  slope.  The single free parameter  of the ENS
model only  controls the normalization  of the $c-\Mvir$  relation, so
that their model  cannot be tuned to fit the data.   The ENS model has
also  been shown  to be  unable to  match the  slope of  the $c-\Mvir$
relation for low mass haloes  at $z=3$ (Col{\'{\i}}n \etal 2004).  For
the Bullock \etal  (2001a) model our data is well  fitted with a model
with  $F=0.01$ and  $K=3.4  \pm  0.1$ (where  the  error reflects  our
estimate  of cosmic  variance).  Note  that this  normalization  is 15
percent lower  than the $K=4.0$  originally proposed by  Bullock \etal
(2001a), but it is in good agreement with Zhao \etal (2003) and Kuhlen
\etal  (2005), who  found a  best-fit normalization  of  $K=3.5$. This
discrepancy is at least partially due to an inconsistency with the use
of transfer functions  in the work of Bullock et al.   If they use the
same  transfer functions  to set  up the  initial conditions  of their
simulations  and  to  compute   the  model  predictions,  they  obtain
$K=3.75$.  This is however  significantly higher than can be accounted
for by our estimate of  cosmic variance.  We find an intrinsic scatter
in $c$  and $\lamp$ at  fixed $\Mvir$ of  $\sigma_{\ln c}=0.33\pm0.03$
and $\sigma_{\ln  \lam}=0.55\pm0.01$, and  a median spin  parameter of
$\lamp=0.034\pm0.001$,  all  in  good  agreement  with  Bullock  \etal
(2001a,b).

In an attempt  to distinguish between relaxed and  unrelaxed haloes we
introduce a new and simple  parameter: $\xoff$ which is defined as the
distance between  the most bound particle  and the center  of mass, in
units  of   the  virial  radius.   The  distribution   of  $\xoff$  is
approximately log-normal  with a median  $\overline{x}_{\rm off}\simeq
0.04$.   The full  set  of haloes  shows  strong correlations  between
$\xoff$  and   the  residuals  of  the   $c-\Mvir$  and  $\lamp-\Mvir$
relations, such  that haloes  with larger $\xoff$  have a  larger than
average $\lamp$  and a lower  than average $c$.  Removing  haloes with
large  $\xoff$ therefore  results in  a higher  mean  concentration, a
lower mean spin  parameter, and in less scatter  in both the $c-\Mvir$
and $\lamp-\Mvir$  relations.  The median spin  parameter of `relaxed'
(GOOD) haloes  is $\lamp=0.030\pm0.003$  with an intrinsic  scatter of
$\sigma_{\ln\lamp}=0.52\pm0.01$.  The  average $c(\Mvir)$ of `relaxed'
haloes is  again in  good agreement  with the model  of B01,  but with
$F=0.01$  and  $K=3.7$,  and  with  a  reduced  intrinsic  scatter  of
$\sigma_{\ln c}=0.26\pm0.03$.  

A better  fit to the $c-\Mvir$  relation for high  mass haloes ($\Mvir
\gta 10^{13} \hMsun$) can be  obtained with $F=0.001$ and $K=2.6$ (for
all haloes), and  $K=2.9$ (for GOOD haloes). This is, however, at the
expense of under predicting the concentrations for the low mass haloes
($\Mvir \lta 10^{11} \hMsun$).

We also find  a strong correlation between the mean  axis ratio of the
haloes,  $\qbar$, and $\xoff$,  such that  more prolate  haloes (i.e.,
those  with lower  $\qbar$)  have higher  $\xoff$,  on average.   This
suggests  that the  majority  of  haloes with  small  axis ratios  are
unrelaxed, and thus  that fragile LSB galaxies are  unlikely to reside
in haloes  that are  strongly aspherical.  This  makes it  less likely
that the discrepancy between observed and predicted rotation curves is
due to  the fact that disks  are strongly elliptical,  as suggested by
Hayashi \etal (2004).
 
We have also investigated  the environment dependence of the residuals
in $c$  and $\lamp$.  This  is interesting because disk  galaxies, and
LSBs   in  particular,   are  preferentially   found  in   regions  of
intermediate to low density (Mo \etal 1994; Rosenbaum \& Bomans 2004).
Defining  `environment' by  the matter  overdensity within  spheres of
radii $1,2,4$ and $8h^{-1}$Mpc, we  find at the low mass end ($M<M^*$)
that more  concentrated haloes live in denser  environments than their
less concentrated  counterparts of the  same mass. This  is consistent
with the studies of Wechsler \etal  (2006) and Gao \& White (2007) who
found a similar trend using correlation functions. However, as we have
shown, this trend is weak  compared to the scatter, which explains why
Lemson \&  Kauffmann (1999) did not notice  any environment dependence
in their  simulation.  Contrary to  the halo concentrations,  the halo
spin parameters  reveal no environment dependence at  fixed mass. This
is at  odds with  the results of  Gao \&  White (2007) who  found that
haloes with  a large spin  parameter are more strongly  clustered than
low spin haloes  of the same mass.  Finally we find  a weak trend that
the most spherical haloes reside in slightly denser environments.

Finally,  using   a  simple  model  for  disk   galaxy  formation,  we
investigated the properties of the  (expected) host haloes of LSB disk
galaxies  (i.e., those  with  a central  surface  brightness $24  \gta
\mu_{0,\rm B}  \gta 23 \,{\rm magn.} {\rm  arcsec}^{-2}$). In addition
to  having higher  than  average spin  parameters,  in agreement  with
numerous  other studies  (e.g.,  Dalcanton \etal  1997; Jimenez  \etal
1998),  we  also  find that  the  host  haloes  of LSB  galaxies  have
concentrations  that  are  biased   low  by  about  15  percent.  This
correlation between halo concentration and disk surface brightness (or
density) owes largely to a (weak) anti-correlation between $\lamp$ and
$c$, and to  the fact that centrifugal equilibrium  commands that less
concentrated haloes  host less  concentrated disks.  The  amplitude of
this correlation is significantly smaller than what has been advocated
by Bailin \etal  (2005), but this owes to the  fact that these authors
did not  remove unrelaxed haloes,  which are unlikely to  host fragile
LSB galaxies, from their analysis.

All these  results have important implications  for the interpretation
of the halo concentrations inferred from LSB rotation curves. Numerous
studies in the past have argued that these are too low compared to the
predictions  for  a $\Lambda$CDM  cosmology  (e.g.,  Alam \etal  2002;
McGaugh \etal  2003; de Blok  \etal 2003). However, there  are several
reasons why we now believe  that the model predictions where too high.
First of all, virtually all  previous predictions were made for a flat
$\Lambda$CDM  cosmology  with  $\Omega_m=0.3$ and  $\sigma_8=0.9$  (or
$\sigma_8=1.0$).  However, if one  adopts the cosmology favored by the
three year data release of  the WMAP mission (Spergel \etal 2006), one
predicts  concentrations that  are  about 25  percent lower  (Macci\`o
\etal,  in  preparation).   Compared  to  a  $\Lambda$CDM  model  with
$\Omega_m=0.3$  and $\sigma_8=1.0$ the  concentrations are  35 percent
lower.   Secondly,   the  B01   model  (with  $F=0.01$   and  $K=4.0$)
overpredicts the halo concentrations by $\sim 15$ percent.  If we take
into account that LSBs only  reside in relaxed haloes, this is lowered
to a $\sim  8$ percent effect.  And finally, one  needs to correct for
the fact that LSB galaxies reside in a biased sub-set of haloes, which
is  another 15  percent effect.   Combining all  these effects,  it is
clear that the halo concentrations  predicted were almost a factor two
too large.  This brings models and data into much better agreement.

\section*{Acknowledgments} 

We  are grateful  to  James  Bullock and  Andrew  Zentner for  sharing
information regarding the normalization of the halo concentrations. 
We also thank an anonymous referee for useful comments that improve the 
presentation of the paper.
A.V.M thanks Justin Read for useful discussions  during the preparation
of this work.  A.A.D. has  been partly supported by the Swiss National
Science Foundation  (SNF).  All the N-body  numerical simulations were
performed          on         the          zBox2         supercomputer
(http://www-theorie.physik.unizh.ch/$\sim$dpotter/zbox2/)    at    the
University of Z\"urich.


\label{lastpage}

\clearpage

\clearpage

\clearpage

\clearpage

\clearpage

\clearpage

\clearpage

\clearpage

\end{document}